\documentclass[11pt]{article}

\usepackage[T1]{fontenc}
\usepackage{lmodern}
\usepackage[utf8]{inputenc}
\usepackage{microtype}
\usepackage{framed}
\usepackage{listings}
\usepackage{vmargin}
\usepackage{setspace}
\usepackage{mathrsfs, mathenv}
\usepackage{amsmath, amsthm, amssymb, amsfonts, amscd}
\usepackage{graphicx}
\usepackage{epstopdf}
\usepackage[svgnames]{xcolor}
\usepackage{hyperref}
\hypersetup{citecolor=blue, colorlinks=true, linkcolor=black}
\usepackage[ruled, vlined]{algorithm2e}
\usepackage[capitalise]{cleveref}
\usepackage{subcaption}
\usepackage{bbm}
\usepackage{cite}
\usepackage{verbatim}
\usepackage{pgfplots}
\usepackage{tikz}
\usetikzlibrary{arrows,decorations.pathmorphing,backgrounds,positioning,fit,matrix}
\usepackage{etoolbox}
\usepackage{color}
\usepackage{lipsum}
\usepackage{ifthen}
\usepackage[title]{appendix}
\usepackage{autonum}
\usepackage{accents}
\usepackage{xpatch}
\usepackage[]{algpseudocode}
\usepackage{multicol}
\usepackage[abs]{overpic}
\usepackage{eqnarray}
\usepackage{bm}

\crefname{assumption}{Assumption}{Assumptions}

\theoremstyle{plain}
\newtheorem{theorem}{Theorem}[section]

\newtheorem{proposition}[theorem]{Proposition}
\numberwithin{equation}{section}

\theoremstyle{definition}

\theoremstyle{remark}
\newtheorem{remark}[theorem]{Remark}

\ifpdf
  \DeclareGraphicsExtensions{.eps,.pdf,.png,.jpg}
\else
  \DeclareGraphicsExtensions{.eps}
\fi

\usepackage{mathtools}

\usepackage{booktabs}

\usepackage{pgfplots}
\usepackage{tikz}
\usetikzlibrary{patterns,arrows,decorations.pathmorphing,backgrounds,positioning,fit,matrix}
\usepackage[labelfont=bf]{caption}
\usepackage{here}
\usepackage[font=normal]{subcaption}

\usepackage{enumitem}
\setlist[itemize]{leftmargin=.5in}
\setlist[enumerate]{leftmargin=.5in,topsep=3pt,itemsep=3pt,label=(\roman*)}

\newcommand*\samethanks[1][\value{footnote}]{\footnotemark[#1]}

\newcommand{\TheTitle}{Improved multifidelity Monte Carlo estimators based on normalizing flows and dimensionality reduction techniques}
\newcommand{\TheAuthors}{A. Zanoni, G. Geraci, M. Salvador, K. Menon, A. L. Marsden, D. E. Schiavazzi}
\title{\TheTitle}
\author{
Andrea Zanoni \thanks{Institute for Computational and Mathematical Engineering, Stanford University, Stanford, CA, USA.}
\and Gianluca Geraci \thanks{Center for Computing Research, Sandia National Laboratories, Albuquerque, NM, USA.}
\and Matteo Salvador \samethanks[1]
\and Karthik Menon \samethanks[1]
\and Alison L. Marsden \samethanks[1] \thanks{Pediatric Cardiology, Stanford University, Stanford, CA, USA.} \thanks{Bioengineering, Stanford University, Stanford, CA, USA.}
\and Daniele E. Schiavazzi \thanks{Department of Applied and Computational Mathematics and Statistics, University of Notre Dame, Notre Dame, IN, USA.}
}
\date{}

\newcommand{\LF}{\mathrm{LF}}
\newcommand{\HF}{\mathrm{HF}}
\newcommand{\MC}{\mathrm{MC}}
\newcommand{\A}{\mathrm{A*}}
\newcommand{\AS}{\mathrm{AS}}
\newcommand{\NF}{\mathrm{NF}}
\newcommand{\NN}{\mathrm{NN}}
\renewcommand{\AE}{\mathrm{AE}}
\newcommand{\MFMC}{\mathrm{MFMC}}
\DeclareMathOperator{\erf}{erf}

\usepackage{amsopn}

\newcommand{\abs}[1]{\left\lvert#1\right\rvert}
\newcommand{\norm}[1]{\left\|#1\right\|}
\renewcommand{\Pr}{\mathbb{P}}

\newcommand{\R}{\mathbb{R}}

\newcommand{\Var}{\operatorname{\mathbb{V}ar}}
\newcommand{\Cov}{\operatorname{\mathbb{C}ov}}
\newcommand{\E}{\operatorname{\mathbb{E}}}

\newcommand{\dd}{\,\mathrm{d}}
\definecolor{shade}{RGB}{100, 100, 100}
\definecolor{bordeaux}{RGB}{128, 0, 50}

\usepackage[usestackEOL]{stackengine}

\definecolor{leg1}{RGB}{0,114,189}
\definecolor{leg2}{RGB}{217,83,25}
\definecolor{leg3}{RGB}{237,177,32}
\definecolor{leg4}{RGB}{126,47,142}
\definecolor{leg5}{RGB}{119,172,48}

\definecolor{leg21}{RGB}{62,38,169}
\definecolor{leg22}{RGB}{46,135,247}
\definecolor{leg23}{RGB}{55,200,151}
\definecolor{leg24}{RGB}{254,195,56}

\ifpdf
\hypersetup{
	pdftitle={\TheTitle},
	pdfauthor={\TheAuthors}
}
\fi

\begin{document}
	
\maketitle	

\begin{abstract} 
\noindent We study the problem of multifidelity uncertainty propagation for computationally expensive models. In particular, we consider the general setting where the high-fidelity and low-fidelity models have a dissimilar parameterization both in terms of number of random inputs and their probability distributions, which can be either known in closed form or provided through samples. We derive novel multifidelity Monte Carlo estimators which rely on a shared subspace between the high-fidelity and low-fidelity models where the parameters follow the same probability distribution, i.e., a standard Gaussian. We build the shared space employing normalizing flows to map different probability distributions into a common one, together with linear and nonlinear dimensionality reduction techniques, active subspaces and autoencoders, respectively, which capture the subspaces where the models vary the most. We then compose the existing low-fidelity model with these transformations and construct modified models with an increased correlation with the high-fidelity model, which therefore yield multifidelity estimators with reduced variance. A series of numerical experiments illustrate the properties and advantages of our approaches.
\end{abstract}


\section{Introduction}

Uncertainty quantification has become a crucial component of computational modeling, as a way to enhance the validity and utility of numerical simulations. Uncertainty quantification studies can provide confidence metrics for quantities of interest and inform future data collection through sensitivity and identifiability analysis. However, in many cases, a naive approach to uncertainty quantification quickly becomes computationally infeasible as a result of the large computational cost needed to numerically solve complex physics-based mathematical models. Therefore, maintaining a reasonable computational cost for an uncertainty quantification study becomes challenging when relying solely on high-fidelity simulations. This has motivated the development of multilevel and multifidelity Monte Carlo strategies that offset the computational cost of estimation to low-fidelity models, accelerating convergence and improving the tractability of uncertainty quantification for computationally expensive simulations \cite{Gil08,NgW14,PWG16,GGE20,Bomarito2022JCP,Schaden_U_SISC_2020,Schaden2021,Croci2023}.

In some cases, however, the low fidelity-model can be obtained through a substantial simplification of the high-fidelity model, resulting in input parameters with potentially both different dimensionality and probability distribution, which we refer to as dissimilar parameterization. We remark that in this case the performance of standard multifidelity Monte Carlo estimators decreases, and it is important to concentrate the variability in few dimensions to enhance the correlation \cite{GEG18,ZGE23}. We therefore propose two methodologies to create a shared subspace of reduced dimension which acts as a bridge between the two models. The first step in obtaining the shared subspace consists of finding an ``important'' subspace individually for both the high-fidelity and low-fidelity models where they vary the most, and consequently which captures most of the variance of the models. Perturbations of the random inputs along such important directions are responsible for most of the variability in the model response. If such a structure is present in the problem, then we can consider model responses only for inputs in the subspace, effectively reducing the problem dimensionality. In this work we employ two different dimensionality reduction techniques, active subspaces and autoencoders, which provide linear and nonlinear transformations, respectively. Active subspaces have been first formalized in \cite{CWD11,CDW14}, while an introduction about autoencoder for unsupervised learning can be found in~\cite{Bal12}. Autoencoders have become popular in the past few years thanks to their expressibility, as they leverage neural networks to find nonlinear transformations of the data, as opposed to linear maps provided by active subspaces. The second ingredient to obtain a shared subspace is a normalizing flow, i.e., an invertible transformation from a generic probability distribution into an easy-to-sample base distribution, usually a standard Gaussian. For a comprehensive review about normalizing flows we refer to~\cite{KPB21,PNR21}. The goal of normalizing flows is enforcing the same probability distribution, in particular a standard Gaussian, of the latent variables, i.e., the parameters in the shared space. For active subspaces, which are known to preserve Gaussianity, we build normalizing flows from the input distributions of the parameters, such that the reduced subspaces of both the high-fidelity and low-fidelity models are automatically Gaussian. On the other hand, for autoencoders we learn normalizing flows from the distributions of the latent variables of the model into the shared subspace, which is therefore Gaussian. We remark that these subspaces and normalizing flows are approximated from the results of a pilot run that is typically employed as a first step in any multifidelity estimator, without the need to run additional high-fidelity simulations.

We reasonably assume that if the variability of the high-fidelity and low-fidelity models is concentrated in few variables, then using the shared space as a bridge aligning the important directions of the fidelities would improve the correlations and consequently reduce the variance of the multifidelity estimator. Starting from the existing low-fidelity model, we therefore construct new low-fidelity models whose inputs are the same parameters of the high-fidelity model which are transformed into inputs for the original low-fidelity model in such a way that the correlation between the fidelities increases. We finally obtain multifidelity estimators which are unbiased, as the high-fidelity model does not change, and with reduced variance with respect to standard multifidelity Monte Carlo estimators. The present work generalizes and extends a series of contributions in this area like, e.g., \cite{GeE18,GEG18,ZGE23}, where the challenge of dissimilar parameterization has been tackled with either active subspaces or adaptive basis \cite{TiG14}. The main extension regarding previous work on active subspaces is that we include a normalizing flow which allows for any input distribution of the parameters, even known only through samples. Moreover, both active subspaces and adaptive basis are linear dimension reduction strategies, and therefore autoencoders are a natural extension since they provide nonlinear transformations.

We apply our methodology to challenging examples, such as reaction-diffusion equations with applications in biological pattern formation and cardiovascular simulations for a coronary model with stenosis. Uncertainty quantification has recently gained momentum in the field of cardiovascular modeling, with recent works exploring a range of methods which can be used to address various sources of uncertainty within these models~\cite{EDS15,StM18,SSK20,SFK20,FGS20,SRD23}. Moreover, various simplifying assumptions can be made to cardiovascular hemodynamics to generate low-fidelity models of intermediate complexity for multifidelity uncertainty propagation. In fact, integrating the Navier--Stokes equations on the vessel cross sections leads to one-dimensional hemodynamic models \cite{HuL73}, and a linearization of the incompressible Navier--Stokes equations around rest conditions leads to an even simpler zero-dimensional formulation utilizing analogous electrical circuits to solve vascular networks \cite{QRV01,RSA22}. These low fidelity models grant us multiple orders of magnitude cost savings over a full three-dimensional model.

\paragraph{Outline.} The rest of the paper is organized as follows. In \cref{sec:current_methods} we give background on multifidelity uncertainty quantification, active subspaces, autoencoders, and normalizing flows, which we employ in the definition of our novel methods, which are described in details in \cref{sec:new_methods}. Then, in \cref{sec:numerical_experiments} we present several numerical experiments, and we finally draw conclusions in \cref{sec:conclusion}.

\section{Review of current methods} \label{sec:current_methods}

In this section we briefly review the main tools employed in our uncertainty quantification pipelines. We recall that our goal is the efficient estimation of scalar quantities of interest of computationally expensive models. Let $Q \colon \R^d \to \R$ represent a computational model, and let $\bm \xi \in \R^d$ be a random vector of inputs distributed according to the joint distribution $\mu$. We aim to characterize the statistical moments of the quantity of interest $Q(\bm \xi)$, focusing in particular on its expectation $\E [Q(\bm \xi)]$. For higher-order moments, we can just replace the quantity of interest $Q(\bm \xi)$ by its power $Q^m(\bm \xi)$ with $m > 1$. The standard Monte Carlo estimator approximates this expected value through a set of $N$ realizations $\{\bm \xi_n \}_{n=1}^N$ of the input variable $\bm \xi \sim \mu$ as
\begin{equation}
\widehat Q^\MC_N = \frac1N \sum_{n=1}^N Q(\bm \xi_n).
\end{equation}
This estimator is simple to compute and it is unbiased, in the sense that 
\begin{equation}
\E \left[ \widehat Q^\MC_N \right] = \E [Q(\bm \xi)].
\end{equation}
However, its root mean squared error is $\mathcal O(N^{-1/2})$, meaning that a large number of evaluations might be necessary in order to reach the desired accuracy. In concrete applications where the evaluation of $Q$ is computationally expensive, increasing the number of samples $N$ can be intractable. Therefore, multifidelity Monte Carlo estimators have been developed to overcome this issue. 

\subsection{Multifidelity Monte Carlo estimator} \label{sec:MFMC}

Let us assume that a computationally cheap model for the original model $Q$ is available, and denote $Q^\HF$ and $Q^\LF$ the original (high-fidelity) and the cheap (low-fidelity) models, respectively. We notice that the low-fidelity model can be any, even biased, approximation of the high-fidelity model, as long as it is computationally cheap to simulate. We adopt the estimator originally introduced in \cite{NgW14}, which focus on the case of a single low-fidelity model, but the multifidelity Monte Carlo estimator can be easily extended, or generalized, to the case of multiple low-fidelity models, e.g., ~\cite{PWG16,GGE20,Bomarito2022JCP,Schaden_U_SISC_2020,Schaden2021,Croci2023}. Let $w = \mathcal C^\LF / \mathcal C^\HF$ be the cost ratio between the two fidelities, and let $\mathcal B$ be the computational budget available in terms of high-fidelity evaluations, i.e.,
\begin{equation}
\mathcal B = N^\HF + w N^\LF,
\end{equation} 
where $N^\HF$ and $N^\LF$ are the numbers of high-fidelity and low-fidelity evaluations, respectively. We aim to split the computationally budget between the high-fidelity and low-fidelity models in such a way that the final multifidelity estimator has the smallest possible variance \cite{PWG16}. Assuming that our budget is large enough, this is achieved by setting
\begin{equation} \label{eq:optimal_allocation}
N^\HF = \frac{\mathcal B}{1 + w \gamma} \qquad \text{and} \qquad N^\LF = \gamma N^{\HF} = \frac{\gamma \mathcal B}{1 + w \gamma}, \qquad \text{with} \qquad \gamma = \sqrt{\frac{\rho^2}{w(1 - \rho^2)}},
\end{equation}
where $\rho$ is the Pearson correlation coefficient between the HF and LF models 
\begin{equation}
\rho = \frac{\Cov \left( Q^{\HF}(\bm \xi), Q^{\LF}(\bm \xi) \right)}{\sqrt{\Var[Q^{\HF}(\bm \xi)] \Var \left [Q^{\LF}(\bm \xi) \right]}}.
\end{equation}
Once the number of evaluations for each model has been selected, the multifidelity Monte Carlo estimator (MFMC) is defined as
\begin{equation} \label{eq:MFMC}
\begin{aligned}
\widehat Q_{N^\HF, N^\LF}^{\MFMC} &= \widehat Q_{N^\HF}^{\HF, \MC} - \beta \left( \widehat Q_{N^\HF}^{\LF, \MC} - \widehat Q_{N^\LF}^{\LF, \MC} \right) \\
&= \frac1{N^\HF} \sum_{n=1}^{N^\HF} Q^\HF(\bm \xi_n) - \beta \left( \frac1{N^\HF} \sum_{n=1}^{N^\HF} Q^\LF(\bm \xi_n) - \frac1{N^\LF} \sum_{n=1}^{N^\LF} Q^\LF(\bm \xi_n) \right),
\end{aligned}
\end{equation}
where the optimal value for the coefficient $\beta$ is given by
\begin{equation} \label{eq:optimal_coefficient}
\beta = \frac{\Cov \left( Q^{\HF}(\bm \xi), Q^{\LF}(\bm \xi) \right)}{\Var \left[ Q^{\LF}(\bm \xi) \right]},
\end{equation}
and $N^\HF$ samples are shared between the two models. 

\begin{remark} \label{rem:optimal_variance}
Computing the optimal values for the numbers $N^\HF$ and $N^\LF$ of evaluations and the coefficient $\beta$ in equations \eqref{eq:optimal_allocation} and \eqref{eq:optimal_coefficient}, respectively, is important to take full advantage of the method and get the smallest possible variance for the multifidelity Monte Carlo estimator. However, it is not essential to employ the optimal values for $N^\HF, N^\LF$, and $\beta$, and any choice would still produce an unbiased estimator. In case the optimal values are used, then we obtain
\begin{equation} \label{eq:variance_MFMC}
\Var \left[ \widehat{Q}_{N^\HF, N^\LF}^{\MFMC} \right] = \Var \left[ \widehat{Q}_{\mathcal B}^{\HF,\MC} \right] \left( \sqrt{1 - \rho^2} + \sqrt{w\rho^2} \right)^2,
\end{equation}
which yields that
\begin{equation}
\abs{\rho} > \frac{4w}{(1+w)^2} \qquad \implies \qquad \Var \left[ \widehat{Q}_{N^\HF, N^\LF}^{\MFMC} \right] < \Var \left[ \widehat{Q}_{\mathcal B}^{\HF,\MC} \right].
\end{equation}
Therefore, variance reduction with respect to standard Monte Carlo is guaranteed as long as the HF and LF models are well correlated. 
\end{remark}

This estimator, as with others available in literature, can be obtained as instances of the so-called Approximate Control Variate (ACV) approach introduced in~\cite{GGE20}. As a consequence, the methodologies developed here will be applicable to a larger set of estimators. In the presence of dissimilar parameterization, retaining high correlation among models is paramount for the efficiency of the estimator. In the next section we introduce active subspaces as a way to increase the correlation between the high-fidelity and low-fidelity models. Moreover, active subspaces act as a bridge between models having a different number of inputs. 

\subsection{Active subspaces} \label{sec:AS}

The active subspaces approach is a methodology which is usually employed in uncertainty quantification studies in order to reduce the dimension of the random inputs, without sacrificing accuracy in approximating a quantity of interest \cite{CDW14}. It allows one to find the dominant directions, i.e., the linear subspace where the quantity of interest $Q$ varies the most. Let $C$ be the matrix which quantifies the variation defined as
\begin{equation} \label{eq:C_AS}
C = \E \left[ \nabla Q(\bm \xi) \nabla Q(\bm \xi)^\top \right],
\end{equation}
where the expectation is taken with respect to the measure $\mu$ and the gradient is computed with respect to the variable $\bm \xi$, and which can be approximated using a collection of samples $\{ \bm \xi_n \}_{n=1}^N$ as
\begin{equation}
C \simeq \widetilde C = \frac1N \sum_{n=1}^N \nabla Q(\bm \xi_n) \nabla Q(\bm \xi_n)^\top.
\end{equation}
We remark that the gradient of the model $Q$ is required in order to compute the matrix $C$. It can be approximated through finite differences, linear approximations, surrogate models, or any other technique to compute derivatives numerically. Additional details about how we deal with the gradient in our work are given in \cref{rem:surrogate_AS}. Moreover, we note that more sophisticated strategies could be introduced to reduce the data requirment of this step, especially for the high-fidelity model. For instance, the approximation of $C$ could be obtained via a multifidelity approach as demonstrated in~\cite{Lam2020}. Note that since $\widetilde C$ is a symmetric positive semidefinite matrix, then it admits a real eigenvalue decomposition 
\begin{equation}
\widetilde C = W \Lambda W^\top \qquad \text{with} \qquad W, \Lambda \in \R^{d \times d},
\end{equation}
where $W$ is orthogonal and its columns are the eigenvectors of $C$, and $\Lambda$ is a diagonal matrix which contains the corresponding eigenvalues $\lambda_1 \ge \lambda_2 \ge \dots \ge \lambda_d \ge 0$, which can be arranged in decreasing order. This ordering suggests a possible separation between important (or active) and irrelevant (or inactive) contributions of the corresponding eigenvector to the linear decomposition of $C$ as a sum of rank-one matrices, in particular if the smallest eigenvalues are close to zero. To better visualize such separation, we write
\begin{equation}
\Lambda = 
\begin{bmatrix}
\Lambda_A & \\ 
& \Lambda_I
\end{bmatrix} \qquad \text{and} \qquad 
W = 
\begin{bmatrix}
W_A & W_I
\end{bmatrix},
\end{equation}
where $\Lambda_A \in \R^{r \times r}$ and $W_A \in \R^{d \times r}$ contain the first $r < d$ eigenvalues and eigenvectors, respectively. The column spans of $W_A$ and $W_I$ represent the active and inactive subspace, respectively. These linear transformations allow us to decompose the original input $\bm \xi \in \R^d$ into the active and inactive parts as follows
\begin{equation} \label{eq:decomposition_AS}
\bm \xi = W_A \bm \xi_A + W_I \bm \xi_I, \qquad \text{where} \qquad \bm \xi_A = W_A^\top \bm \xi \in \R^r, \quad \xi_I = W_I^\top \bm \xi \in \R^{d-r}.
\end{equation}
If the dimensionality of the problem can actually be reduced, then the inactive component of the decomposition can be ignored because it gives a negligible contribution to the quantity of interest, i.e., $Q(\bm \xi) \simeq Q(W_A \bm \xi_A)$, and the problem becomes $r$-dimensional.

\begin{remark}
The probability distribution of the random inputs $\mu$ often results from a modeling choice, particularly when observational data are insufficient, and the matrix $C$ in equation~\eqref{eq:C_AS} is dependent on this distribution. Therefore, we notice that different choices for $\mu$ lead to different active subspaces for the same model.
\end{remark}

The distribution of the active part in the decomposition \eqref{eq:decomposition_AS} is in general unknown. Nevertheless, if the distribution of the data $\mu$ is the standard Gaussian, i.e., $\bm \xi \sim \mathcal N(\bm 0, I_d)$, then the active component remains standard Gaussian, i.e., $\bm \xi_A \sim \mathcal N(\bm 0, I_r)$, and in fact a linear transformation of a Gaussian random variable is still Gaussian, and it holds
\begin{equation}
\E [\bm\xi_A] = W_A^\top \E [\bm\xi] = 0 \qquad \text{and} \qquad \Var [\bm\xi_A] = W_A^\top \Var [\bm\xi] W_A = W_A^\top W_A = I_r.
\end{equation}
This suggests the possibility of generating shared parameterizations even for models having a different number of random inputs, as long as their active inputs share the same dimensionality and distribution \cite{GeE18,GEG18}. For general probability distributions, it is often possible to define a transformation mapping $\mu$ to a standard Gaussian via normalizing flows, compute the active subspace, and finally transform the variables back to their original distribution after sampling. We notice that this is a significant advantage since it allows one to handle any type of input data, independently of their correlation. As observed in~\cite{GeE18}, even if a transformation introduces additional complexity, the increase in computational cost could be outweighed by the increase in correlation between models. A strong limitation of the active subspace technique is that it provides only linear maps for dimensionality reduction. This restriction can be overcome by replacing active subspaces with autoencoders, which allow for nonlinear transformations. The next two sections will therefore focus on autoencoders and normalizing flows, respectively.

\subsection{Autoencoders} \label{sec:AE}

Autoencoders are a data-driven approach widely used for unsupervised dimensionality reduction, with the ability to learn an intrinsic structure existing in the data by leveraging neural networks. They consist of one encoder $\mathcal E$ followed by one decoder $\mathcal D$, where the encoder $\mathcal E \colon \R^d \to \R^r$ with $r < d$ compresses the original input $\bm \xi \in \R^d$ into a latent representation $\bm x \in \R^r$, and the decoder $\mathcal D \colon \R^r \to \R^d$ seeks to reconstruct the original input $\bm \xi$ starting from the latent representation $\bm x$, in the sense that $\mathcal D(\mathcal E(\bm \xi)) \simeq \bm \xi$. The functions $\mathcal E$ and $\mathcal D$ are usually parameterized by fully connected neural networks. In this work we are not interested in reconstructing exactly the original input, but rather $Q(\bm \xi)$. In particular, given a model $Q$, analogously to the active subspace approach, we would like to find a lower dimensional representation of $Q$ by selecting a manifold of dimension $r$ where the function varies the most. Hence, we aim to reconstruct the original quantity of interest, meaning that $Q(\mathcal D(\mathcal E(\bm \xi))) \simeq Q(\bm \xi)$. We notice that this can be seen as a supervised dimensionality reduction, where we would like to learn a structure in the data according to some metric, which in this case is the model $Q$. We parameterize the encoder and the decoder as fully connected neural networks $\mathcal E(\cdot; \phi_E)$ and $\mathcal D(\cdot; \phi_D)$ with hyperbolic tangent activation functions, and we compute the optimal parameters by minimizing the loss function
\begin{equation}
\mathcal L_\AE(\phi_E, \phi_D) = \frac1N \sum_{n=1}^N \abs{Q(\bm \xi_n) - Q(\mathcal D(\mathcal E(\bm \xi_n; \phi_E); \phi_D))},
\end{equation}
where $\{ \bm \xi_n \}_{n=1}^N$ is the sample of available data. We notice that minimizing the loss function requires multiple evaluations of the model $Q$, and this can be impractical if the model is computationally expensive. We therefore build a surrogate model which is only used in the training process, as we discuss in \cref{rem:surrogate_AE}. Using the same terminology as for the active subspaces technique, we say that the active variable is $\bm \xi_A = \mathcal E(\bm \xi)$ and we notice that its distribution is unknown. Therefore, by constructing a map from the probability distribution of the active variable into a standard Gaussian through normalizing flows (see \cref{sec:NF}), we can generate a shared parameterization even for models having a different number of inputs, as long as the reduced dimension of the autoencoder is the same.

\subsection{Normalizing flows} \label{sec:NF}

Normalizing flows are invertible transformations which map generic probability distributions into more simple and tractable distributions, usually standard Gaussian. Let $\mu$ be a probability distribution on $\R^d$ with density $\psi$ and let $\mu_0$ with density $\psi_0$ be the target distribution, which in our work, like in most cases, is $\mu_0 = \mathcal N(\bm 0,I_d)$. We aim to find a diffeomorphism $\mathcal T \colon \R^d \to \R^d$ such that $\mathcal T_\# \mu = \mu_0$, where $\mathcal T_\#$ denotes the pushforward measure through the map $\mathcal T$. We consider a parameterization $\mathcal T(\cdot; \theta)$, which is the composition of invertible transformations defined by neural networks. Then, given a sample $\{ \bm \xi_n \}_{n=1}^N$ from the initial distribution $\mu$, the best parameters $\theta$ are computed by maximizing the log-likelihood function for the density $\widetilde \psi(\cdot; \theta)$ of the measure $\mathcal T^{-1}(\cdot; \theta)_\# \mu_0$. In practice, we minimize the loss function 
\begin{equation}
\mathcal L_\NF(\theta) = - \sum_{n=1}^N \log \widetilde \psi(\bm \xi_n; \theta),
\end{equation}
which, by the change of variable formula, can be written as
\begin{equation} \label{eq:loss_NF}
\mathcal L_\NF(\theta) = - \sum_{n=1}^N \left[ \log \psi_0(\mathcal T(\bm \xi_n; \theta)) + \log \abs{\det \nabla \mathcal T(\bm \xi_n; \theta)} \right].
\end{equation}

\begin{remark} \label{rem:loss_NF}
If the target distribution is standard Gaussian, i.e., $\mu_0 = \mathcal N(\bm 0,I_d)$, then $\psi_0$ is the density of a multivariate normal distribution, and equation \eqref{eq:loss_NF} reads
\begin{equation}
\mathcal L_\NF(\theta) = \sum_{n=1}^N \left[ \frac12 \norm{\mathcal T(\bm \xi_n; \theta)}^2 - \log \abs{\det \nabla \mathcal T(\bm \xi_n; \theta)} \right] + \frac{Nd}2 \log(2 \pi),
\end{equation}
where the last term in the right-hand side can be neglected since it is independent of $\theta$.
\end{remark}

Therefore, a good parameterization for a normalizing flow needs to be sufficiently expressive, in order to be able to approximate the exact transformation, and efficient in terms of computation of the map itself, and the determinant of its Jacobian matrix. Moreover, the computation of the inverse of the map should also be cheap, so that we can draw new samples from the initial distribution $\mu$ efficiently. Indeed, a straightforward way to get a new sample from $\mu$ consists of drawing a sample $\bm x \sim \mu_0$ and then applying the inverse of the learned map, i.e., $\bm \xi = \mathcal T^{-1}(\bm x; \theta)$. In this work we employ both the RealNVP normalizing flow, which is presented in \cite{DSB17} and implemented in \cite{SLC23}, and splines from the FlowTorch package (\url{https://www.flowtorch.ai}). We also include a deterministic transformation (inverse of the hyperbolic tangent) to map a distribution with finite support, e.g., uniform, into a distribution with infinite support, e.g., standard Gaussian. We remark that normalizing flows have already been used in uncertainty quantification. Some examples are \cite{MMP16,SBM18}, where a specific type of map given by the the Knothe-Rosenblatt rearrangement is employed, and \cite{WLS22}, where normalizing flows are combined with adaptive surrogate modeling.

\section{Enhancing multifidelity estimator performance} \label{sec:new_methods}

\begin{figure}[t]
\centering
\includegraphics[scale=0.45]{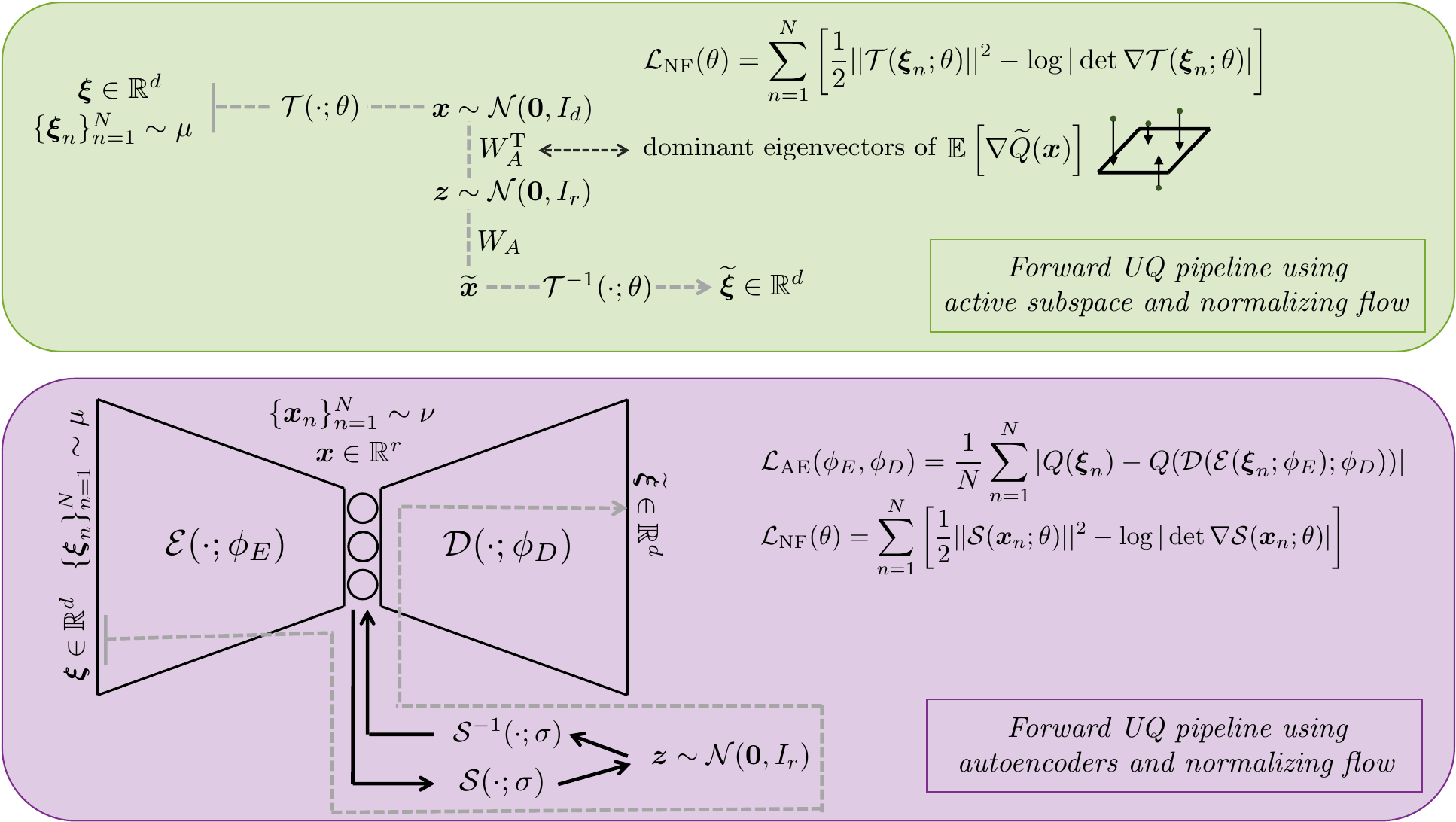}
\caption{Summary of the methodologies presented in \cref{sec:new_methods}, which holds for both the high-fidelity and low-fidelity models.}
\label{fig:methods}
\end{figure}

We now describe our two methodologies for improving multifidelity Monte Carlo estimators. Our pipelines have the twofold purpose of reducing the variance of the estimator and increasing its range of applicability to problems where the original model and its low-fidelity version have a dissimilar parameterization. We recall that we consider an expensive high-fidelity model $Q^\HF \colon \R^{d^\HF} \to \R$ and a cheap low-fidelity model $Q^\LF \colon \R^{d^\LF} \to \R$, and we aim to estimate the expectation of the former model $\E[Q^\HF(\bm \xi^\HF)]$ under the assumption that the input parameters $\bm \xi^\HF \in \R^{d^\HF}$ and $\bm \xi^\LF \in \R^{d^\LF}$ are distributed according to some probability distribution $\mu^\HF$ and $\mu^\LF$, respectively. We notice that, in the most general setting, we allow the number of input parameters $d^\HF,d^\LF$ and also their distributions $\mu^\HF, \mu^\LF$ to be different. Our goal is to create a shared parameterization of reduced dimension $r$, where the two models vary the most. We remark that we do not require the high-fidelity and low-fidelity models to naturally share the same lower dimensional manifold, but our main goal is finding the lower dimensional subspace of each model separately, and then create a link between them through a shared subspace. We then assume that we want to keep the original high-fidelity model for two reasons. First, we could not have additional resources for generating new high-fidelity simulations on the shared space, and second we would like to preserve the unbiasedness of the multifidelity estimator. On the other hand, we create a modified low-fidelity model which is better correlated to the high-fidelity one. We achieve this by employing normalizing flows and dimensionality reduction techniques, in particular active subspaces for the method in \cref{sec:MFMC_AS} and autoencoders for the method in \cref{sec:MFMC_AE}. We remark that the main difference between the two approaches is that autoencoders allow for nonlinear transformations which cannot be obtained from the active subspace technique. A summary of the two methodologies, which are outlined in the next two sections, is showed in \cref{fig:methods}.

\subsection{Coupling MFMC with active subspaces and normalizing flows} \label{sec:MFMC_AS}

In this section we create a shared parameterization between $Q^\HF$ and $Q^\LF$ using the active variables given by the active subspaces of the two models. Let $\{ \bm \xi^\HF_n \}_{n=1}^N \sim \mu^\HF$ and $\{ \bm \xi^\LF_n \}_{n=1}^N \sim \mu^\LF$ be samples from the input distributions, and consider the normalizing flows $\mathcal T^\HF(\cdot; \theta^\HF) \colon \R^{d^\HF} \to \R^{d^\HF}$ and $\mathcal T^\LF(\cdot; \theta^\LF) \colon \R^{d^\LF} \to \R^{d^\LF}$ such that
\begin{equation} \label{eq:NF_AS_pushforward}
\mathcal T^\HF(\cdot; \theta^\HF)_\# \mu^\HF = \mathcal N(\bm 0, I_{d^\HF}) \qquad \text{and} \qquad \mathcal T^\LF(\cdot; \theta^\LF)_\# \mu^\LF = \mathcal N(\bm 0, I_{d^\LF}),
\end{equation}
and which, due to \cref{rem:loss_NF}, are obtained minimizing the loss functions
\begin{equation}
\begin{aligned}
\mathcal L_\NF^\HF(\theta^\HF) &= \sum_{n=1}^N \left[ \frac12 \norm{\mathcal T^\HF(\bm \xi^\HF_n; \theta^\HF)}^2 - \log \abs{\det \nabla \mathcal T^\HF(\bm \xi^\HF_n; \theta^\HF)} \right], \\
\mathcal L_\NF^\LF(\theta^\LF) &= \sum_{n=1}^N \left[ \frac12 \norm{\mathcal T^\LF(\bm \xi^\LF_n; \theta^\LF)}^2 - \log \abs{\det \nabla \mathcal T^\LF(\bm \xi^\LF_n; \theta^\LF)} \right].
\end{aligned}
\end{equation}
We can now define the modified models $\widetilde Q^\HF \colon \R^{d^\HF} \to \R$ and $\widetilde Q^\LF \colon \R^{d^\LF} \to \R$, whose input distributions are standard Gaussian, employing the inverse of the normalizing flows
\begin{equation} \label{eq:modified_models_AS}
\widetilde Q^\HF(\bm x^\HF) = Q^\HF((\mathcal T^\HF)^{-1}(\bm x^\HF; \theta^\HF)) \quad \text{and} \quad \widetilde Q^\LF(\bm x^\LF) = Q^\LF((\mathcal T^\LF)^{-1}(\bm x^\LF; \theta^\LF)).
\end{equation}
Applying the procedure described in \cref{sec:AS} to $\widetilde Q^\HF$ and $\widetilde Q^\LF$, we compute the matrices 
\begin{equation} \label{eq:C_HF_LF}
\widetilde C^\HF = \frac1N \sum_{n=1}^N \nabla \widetilde Q^\HF(\bm x^\HF_n) \nabla \widetilde Q^\HF(\bm x^\HF_n)^\top \quad \text{and} \quad \widetilde C^\LF = \frac1N \sum_{n=1}^N \nabla \widetilde Q^\LF(\bm x^\LF_n) \nabla \widetilde Q^\LF(\bm x^\LF_n)^\top,
\end{equation}
where $\bm x^\HF_n = \mathcal T^\HF(\bm \xi^\HF_n; \theta^\HF)$ and $\bm x^\LF_n = \mathcal T^\LF(\bm \xi^\LF_n; \theta^\LF)$, and we learn the active subspaces $W_A^\HF \in \R^{d^\HF \times r}$ and $W_A^\LF \in \R^{d^\LF \times r}$ of dimension $r$, which capture the directions of maximum change of the two models. Hence, the shared space is obtained applying first the normalizing flow and then the transpose of the active subspace matrix as follows
\begin{equation}
\begin{aligned}
\bm z_n^\HF &= (W_A^\HF)^T \bm x^\HF_n = (W_A^\HF)^T \mathcal T^\HF(\bm \xi^\HF_n; \theta^\HF) \in \R^r, \\
\bm z_n^\LF &= (W_A^\LF)^T \bm x^\LF_n = (W_A^\LF)^T \mathcal T^\LF(\bm \xi^\LF_n; \theta^\LF) \in \R^r,
\end{aligned}
\end{equation}
where both $\bm z_n^\HF$ and $\bm z_n^\LF$ are distributed accordingly to a standard Gaussian $\mathcal N(\bm 0, I_r)$. We recall here that $\bm \xi$ stands for the original input, $\bm x$ is the corresponding normally distributed parameter, and $\bm z$ is the active variable in the shared space. We remark that, if the optimal reduced dimensions differ between the two fidelities, the value of $r$ should be chosen in the range given by these two dimensions finding the best trade-off between capturing all the variability of the models and increasing their correlation. We recall that our goal is increasing the correlation between the high-fidelity, which we do not modify, and the low fidelity models. We therefore construct a new low-fidelity model in which we map the input parameters of the high-fidelity model into the input parameters of the original low-fidelity model by means of the shared space. In particular, we define $\mathcal Q_\AS^\LF \colon \R^{d^\HF} \to \R$ as
\begin{equation} \label{eq:LF_AS}
\mathcal Q^\LF_\AS(\bm \xi^\HF) = Q^\LF((\mathcal T^\LF)^{-1}(W_A^\LF (W_A^\HF)^\top \mathcal T^\HF(\bm \xi^\HF; \theta^\HF), \theta^\LF)),
\end{equation}
and we employ it to introduce a new multifidelity estimator
\begin{equation} \label{eq:estimator_MFMCAS}
\widehat Q_{N^\HF, N^\LF}^{\MFMC, \AS} = \frac1{N^\HF} \sum_{n=1}^{N^\HF} Q^\HF(\bm \xi^\HF_n) - \beta \left( \frac1{N^\HF} \sum_{n=1}^{N^\HF} \mathcal Q_\AS^\LF(\bm \xi^\HF_n) - \frac1{N^\LF} \sum_{n=1}^{N^\LF} \mathcal Q_\AS^\LF(\bm \xi^\HF_n) \right).
\end{equation}
We remark that since the high-fidelity model does not change, then $\widehat Q_{N^\HF, N^\LF}^{\MFMC, \AS}$ is unbiased. In \cref{alg:MFMC_AS} we summarize the main steps needed to construct the estimator. The same considerations presented in \cref{sec:MFMC} regarding the choice of the coefficient $\beta$ and of the numbers $N^\HF$ and $N^\LF$ of high-fidelity and low-fidelity evaluations are still valid here. Moreover, we remark that, if we do not have the possibility to generate new high-fidelity simulations, we can fix $N^\HF = N$, where $N$ is the dimension of the pilot sample used to build the new low-fidelity model, and increase only the number $N^\LF$ of low-fidelity simulations. In this case, we would not get the optimal variance reduction outlined in \cref{rem:optimal_variance}, but we can still obtain a reduction in the variance of the multifidelity estimator due to the higher correlation between $Q^\HF$ and $\mathcal Q^\LF_\AS$ with respect to $Q^\HF$ and $Q^\LF$.

\begin{algorithm}
\caption{MFMC AS} \label{alg:MFMC_AS}
\begin{tabbing}
\hspace{-0.25cm}\textbf{Input:} \= High fidelity and low fidelity models $Q^\HF$ and $Q^\LF$ \\
\> Distributions $\mu^\HF$ and $\mu^\LF$ for the input parameters \\
\> \hspace{0.25cm} or samples $\{ \bm \xi^\HF_n \}_{n=1}^N \sim \mu^\HF$ and $\{ \bm \xi^\LF_n \}_{n=1}^N \sim \mu^\LF$ from them \\
\> Computational budget $\mathcal B$.
\end{tabbing}
\begin{tabbing}
\hspace{-0.25cm}\textbf{Output:} \= Estimation $\widehat Q_{N^\HF, N^\LF}^{\MFMC, \AS}$ of $\E[Q^\HF(\bm \xi^\HF)]$.
\end{tabbing}
\begin{enumerate}[label=\arabic*:,itemindent=-0.75cm]
\item Compute the normalizing flows $\mathcal T^\HF(\cdot; \theta^\HF)$ and $\mathcal T^\LF(\cdot; \theta^\LF)$ which satisfy \eqref{eq:NF_AS_pushforward}.
\item Compute the active subspaces $W_A^\HF$ and $W_A^\LF$ from the matrices $\widetilde C^\HF$ and $\widetilde C^\LF$ \\ \hspace{-0.8cm} in \eqref{eq:C_HF_LF} obtained from the modified models $\widetilde Q^\HF$ and $\widetilde Q^\LF$ in \eqref{eq:modified_models_AS}.
\item Define the new low-fidelity model $\mathcal Q_\AS^\LF$ in \eqref{eq:LF_AS}.
\item Compute the optimal allocation $N^\HF, N^\LF$ from a pilot sample.
\item Compute the estimator $\widehat Q_{N^\HF, N^\LF}^{\MFMC, \AS}$ in \eqref{eq:estimator_MFMCAS}.
\end{enumerate}
\end{algorithm}

\begin{remark} \label{rem:surrogate_AS}
In order to learn the active subspace matrices, it is necessary to compute the gradients of the models in equation \eqref{eq:C_HF_LF}, which might not be available or might be too computationally expensive. An approximation of the gradients can be obtained replacing the partial derivatives with finite differences, but this approach can be unfeasible, in particular in high dimensions, even for the low-fidelity model. Therefore, we propose to train surrogate models $Q^\HF_\NN$ and $Q^\LF_\NN$ for $Q^\HF$ and $Q^\LF$ based on fully connected neural networks with ReLU activation functions, and then compute the gradients using automatic differentiation. Given a set of realizations $\{ ( \bm \xi^\HF_n, Q^\HF(\bm \xi^\HF_n) ) \}_{n=1}^N$ and $\{ ( \bm \xi^\LF_n, Q^\LF(\bm \xi^\LF_n) ) \}_{n=1}^N$, we obtain the neural networks $Q^\HF_\NN(\cdot; \alpha^\HF)$ and $Q^\LF_\NN(\cdot; \alpha^\LF)$ minimizing the loss functions
\begin{equation}
\begin{aligned}
\mathcal L_\NN^\HF(\alpha^\HF) &= \frac1N \sum_{n=1}^N \abs{Q^\HF(\bm \xi^\HF_n) - Q_\NN^\HF(\bm \xi^\HF_n; \alpha^\HF)}, \\
\mathcal L_\NN^\LF(\alpha^\LF) &= \frac1N \sum_{n=1}^N \abs{Q^\LF(\bm \xi^\LF_n) - Q_\NN^\LF(\bm \xi^\LF_n; \alpha^\LF)}.
\end{aligned}
\end{equation}
We notice that these approximations might not be accurate if the number $N$ of data points is not sufficiently large. On the other hand, if the surrogate models were suitably accurate, then it would make more sense to consider the neural network surrogate rather than the low fidelity model. We emphasize here that these surrogate models are employed with the only purpose of helping to find low dimensional subspaces where the models vary the most, and should not be used to approximate the models themselves for the evaluations needed to compute the multifidelity estimator. Hence, even if the approximation provided by the neural networks surrogates can be quite poor, and this is a problem for identifying the best lower-dimensional subspaces, it can still be sufficient for capturing lower-dimensional manifolds that increase the correlation between the models.
\end{remark}

\subsection{Coupling MFMC with autoencoders and normalizing flows} \label{sec:MFMC_AE}

A strong limitation of the method presented in the previous section is that it relies on linear dimensionality reduction. In this section, we extend the previous methodology by replacing active subspaces with the supervised autoencoders introduced in \cref{sec:AE}, and therefore admitting nonlinear transformations for dimensionality reduction. Although nonlinear transformations are more expressive, they do not preserve Gaussianity, and consequently the normalizing flows used to map the input distributions into standard Gaussian become redundant. Hence, given samples $\{ \bm \xi^\HF_n \}_{n=1}^N \sim \mu^\HF$ and $\{ \bm \xi^\LF_n \}_{n=1}^N \sim \mu^\LF$ from the input distributions, we first learn the autoencoders with $r$-dimensional latent variables
\begin{equation} \label{eq:AE_def}
\begin{cases}
\mathcal E^\HF(\cdot; \phi_E^\HF) \colon \R^{d^\HF} \to \R^r \\
\mathcal D^\HF(\cdot; \phi_D^\HF) \colon \R^r \to \R^{d^\HF}
\end{cases}
\qquad \text{and} \qquad
\begin{cases}
\mathcal E^\LF(\cdot; \phi_E^\LF) \colon \R^{d^\LF} \to \R^r \\
\mathcal D^\LF(\cdot; \phi_D^\LF) \colon \R^r \to \R^{d^\LF}
\end{cases},
\end{equation}
by minimizing the loss functions
\begin{equation}
\begin{aligned}
\mathcal L_\AE^\HF(\phi_E^\HF, \phi_D^\HF) &= \frac1N \sum_{n=1}^N \abs{Q^\HF(\bm \xi^\HF_n) - Q^\HF(\mathcal D^\HF(\mathcal E^\HF(\bm \xi^\HF_n; \phi_E^\HF); \phi_D^\HF))}, \\
\mathcal L_\AE^\LF(\phi_E^\LF, \phi_D^\LF) &= \frac1N \sum_{n=1}^N \abs{Q^\LF(\bm \xi^\LF_n) - Q^\LF(\mathcal D^\LF(\mathcal E^\LF(\bm \xi^\LF_n; \phi_E^\LF); \phi_D^\LF))}.
\end{aligned}
\end{equation}
The latent variables of the autoencoders 
\begin{equation}
\bm x^\HF_n = \mathcal E^\HF(\bm \xi^\HF_n; \phi_E^\HF) \in \R^r \qquad \text{and} \qquad \bm x^\LF_n = \mathcal E^\LF(\bm \xi^\LF_n; \phi_E^\LF) \in \R^r
\end{equation}
now share the same dimensionality, but not the same probability distribution. Therefore, in order to create a shared space, we construct two normalizing flows which map the distributions of the latent spaces of the two autoencoders into a standard Gaussian. In particular, let $\nu^\HF = \mathcal E^\HF(\cdot; \phi_E^\HF)_\# \mu^\HF$ and $\nu^\LF = \mathcal E^\LF(\cdot; \phi_E^\LF)_\# \mu^\LF$, and consider the normalizing flows $\mathcal S^\HF(\cdot; \sigma^\HF) \colon \R^r \to \R^r$ and $\mathcal S^\LF(\cdot; \sigma^\LF) \colon \R^r \to \R^r$ such that
\begin{equation} \label{eq:NF_AE_pushforward}
\mathcal S^\HF(\cdot; \sigma^\HF)_\# \nu^\HF = \mathcal N(\bm 0, I_r) \qquad \text{and} \qquad \mathcal S^\LF(\cdot; \sigma^\LF)_\# \nu^\LF = \mathcal N(\bm 0, I_r),
\end{equation}
and which are obtained minimizing the loss functions
\begin{equation}
\begin{aligned}
\mathcal L_\NF^\HF(\sigma^\HF) &= \sum_{n=1}^N \left[ \frac12 \norm{\mathcal S^\HF(\bm x^\HF_n; \sigma^\HF)}^2 - \log \abs{\det \nabla \mathcal S^\HF(\bm x^\HF_n; \sigma^\HF)} \right], \\
\mathcal L_\NF^\LF(\sigma^\LF) &= \sum_{n=1}^N \left[ \frac12 \norm{\mathcal S^\LF(\bm x^\LF_n; \sigma^\LF)}^2 - \log \abs{\det \nabla \mathcal S^\LF(\bm x^\LF_n; \sigma^\LF)} \right].
\end{aligned}
\end{equation}

\begin{remark}
The autoencoders and the normalizing flows are trained sequentially, meaning that we first compute the best autoencoders and then train the normalizing flows on the resulting latent spaces, in order to get latent variables distributed as standard Gaussians. We would also like to point out that we could have used variational autoencoders which would certainly result in a more regular latent space, but without any guarantees on the distribution of the latent space to be a standard Gaussian, unlike normalizing flow where this is guaranteed by construction. Moreover, we remark that, in principle, the normalizing flows could be replaced by any other map between the latent spaces of the autoencoders of the high-fidelity and low-fidelity models, and we leave the determination of the most efficient method for this task to subsequent work.
\end{remark}

We therefore obtain a shared space by applying first the encoder and then the normalizing flow as follows
\begin{equation}
\begin{aligned}
\bm z_n^\HF &= \mathcal S^\HF(\bm x^\HF_n; \sigma^\HF) = \mathcal S^\HF(\mathcal E^\HF(\bm \xi^\HF_n; \phi_E^\HF); \sigma^\HF) \in \R^r, \\
\bm z_n^\LF &= \mathcal S^\LF(\bm x^\LF_n; \sigma^\LF) = \mathcal S^\LF(\mathcal E^\LF(\bm \xi^\LF_n; \phi_E^\LF); \sigma^\LF) \in \R^r,
\end{aligned}
\end{equation}
where both $\bm z_n^\HF$ and $\bm z_n^\LF$ are distributed accordingly to a standard Gaussian $\mathcal N(\bm 0, I_r)$. We recall here that $\bm \xi$ stands for the original input, $\bm x$ is the latent variable of the autoencoder, and $\bm z$ is the corresponding normally distributed variable in the shared space. We remark that the autoencoders do not provide a ranking of the variables in the latent dimension according to their variance contributions like the active subspace technique. Hence, if the latent space is multidimensional, we propose to select the best ordering of the components of the latent variable in terms of the estimated correlation. The next steps are analogous to the previous methodology introduced in \cref{sec:MFMC_AS}. In order to increase the correlation between the high-fidelity and low-fidelity models, and without modifying the former, we construct a new low-fidelity model in which we map the input parameters of the high-fidelity model into the input parameters of the original low-fidelity model by means of the shared space. In particular, we define
\begin{equation} \label{eq:LF_AE}
\mathcal Q_\AE^\LF(\bm \xi^\HF) = Q^\LF(\mathcal D^\LF((\mathcal S^\LF)^{-1}(\mathcal S^\HF(\mathcal E^\HF(\bm \xi^\HF; \phi_E^\HF); \sigma^\HF); \sigma^\LF); \phi_D^\LF)),
\end{equation}
which yields another new multifidelity estimator 
\begin{equation} \label{eq:estimator_MFMCAE}
\widehat Q_{N^\HF, N^\LF}^{\MFMC, \AE} = \frac1{N^\HF} \sum_{n=1}^{N^\HF} Q^\HF(\bm \xi^\HF_n) - \beta \left( \frac1{N^\HF} \sum_{n=1}^{N^\HF} \mathcal Q_\AE^\LF(\bm \xi^\HF_n) - \frac1{N^\LF} \sum_{n=1}^{N^\LF} \mathcal Q_\AE^\LF(\bm \xi^\HF_n) \right),
\end{equation}
which is unbiased since we do not modify the high-fidelity model. In \cref{alg:MFMC_AE} we summarize the main steps needed to construct the estimator. The same observations highlighted for the methodology based on active subspaces still hold here. In particular, we notice that, in case we cannot generate new high-fidelity simulations, we can reuse the evaluations employed for training the autoencoder, and change only the low-fidelity simulations. Moreover, without reaching the optimal allocation for $N^\HF$ and $N^\LF$ we would not get the optimal variance reduction stated in \cref{rem:optimal_variance}, but we would still decrease the variance of the multifidelity estimator due to the higher correlation between $Q^\HF$ and $\mathcal Q^\LF_\AE$ with respect to $Q^\HF$ and $Q^\LF$.

\begin{algorithm}
\caption{MFMC AE} \label{alg:MFMC_AE}
\begin{tabbing}
\hspace{-0.25cm}\textbf{Input:} \= High fidelity and low fidelity models $Q^\HF$ and $Q^\LF$ \\
\> Distributions $\mu^\HF$ and $\mu^\LF$ for the input parameters \\
\> \hspace{0.25cm} or samples $\{ \bm \xi^\HF_n \}_{n=1}^N \sim \mu^\HF$ and $\{ \bm \xi^\LF_n \}_{n=1}^N \sim \mu^\LF$ from them \\
\> Computational budget $\mathcal B$.
\end{tabbing}
\begin{tabbing}
\hspace{-0.25cm}\textbf{Output:} \= Estimation $\widehat Q_{N^\HF, N^\LF}^{\MFMC, \AE}$ of $\E[Q^\HF(\bm \xi^\HF)]$.
\end{tabbing}
\begin{enumerate}[label=\arabic*:,itemindent=-0.75cm]
\item Compute the autoencoders $(\mathcal E^\HF(\cdot; \phi_E^\HF), \mathcal D^\HF(\cdot; \phi_D^\HF))$ and \\ \hspace{-0.8cm} $(\mathcal E^\LF(\cdot; \phi_E^\LF), \mathcal D^\LF(\cdot; \phi_D^\LF))$ in \eqref{eq:AE_def}.
\item Compute the normalizing flows $\mathcal S^\HF(\cdot; \sigma^\HF)$ and $\mathcal S^\LF(\cdot; \sigma^\LF)$ which satisfy \eqref{eq:NF_AE_pushforward}.
\item Select the best ordering of the components of the latent variable.
\item Define the new low-fidelity model $\mathcal Q_\AE^\LF$ in \eqref{eq:LF_AE}.
\item Compute the optimal allocation $N^\HF, N^\LF$ from a pilot sample.
\item Compute the estimator $\widehat Q_{N^\HF, N^\LF}^{\MFMC, \AE}$ in \eqref{eq:estimator_MFMCAE}.
\end{enumerate}
\end{algorithm}

\begin{remark} \label{rem:surrogate_AE}
In order to train the autoencoder, it is necessary to evaluate the models $Q^\HF$ and $Q^\LF$ multiple times, and this not only is impossible for the expensive high-fidelity model, but can also be impractical for the cheaper low-fidelity model. Therefore, we propose to train surrogate models $Q_\NN^\HF$ and $Q_\NN^\LF$ for $Q^\HF$ and $Q^\LF$ based on fully connected neural networks with ReLU activation functions, analogously to what we did in \cref{rem:surrogate_AS}. We recall that these surrogate models might not be highly accurate, in particular for small amount of data, and therefore should not be used for model approximation, but only with the aim of finding nonlinear subspaces of reduced dimension during the training of the autoencoders. We also recall that even if these surrogate models are not sufficiently accurate to determine the best lower-dimensional manifolds, they can still capture nonlinear subspaces that increase the correlation.
\end{remark}

\subsubsection{A particular choice for the autoencoder} \label{sec:particular_autoencoder}

In this section we restrict ourselves to the particular case where the encoder is the model itself, i.e., $\mathcal E^\HF = Q^\HF$ and $\mathcal E^\LF = Q^\LF$. Then, the decoders $\mathcal D^\HF$ and $\mathcal D^\LF$ need to satisfy
\begin{equation}
Q^\HF(\bm \xi^\HF) \simeq Q^\HF(\mathcal D^\HF (Q^\HF(\bm \xi^\HF))) \qquad \text{and} \qquad Q^\LF(\bm \xi^\LF) \simeq Q^\LF(\mathcal D^\LF (Q^\LF(\bm \xi^\LF))),
\end{equation}
which imply
\begin{equation} \label{eq:condition_particular_AE}
\bm x^\HF \simeq Q^\HF(\mathcal D^\HF (\bm x^\HF)) \qquad \text{and} \qquad \bm x^\LF \simeq Q^\LF(\mathcal D^\LF (\bm x^\LF)),
\end{equation}
for all $\bm x^\HF$ in the image of $Q^\HF$ and $\bm x^\LF$ in the image of $Q^\LF$. It is possible to show that the equalities in \eqref{eq:condition_particular_AE} can be satisfied exactly by choosing $\mathcal D^\HF$ and $\mathcal D^\LF$ to be the right inverses of the functions $Q^\HF$ and $Q^\LF$, respectively. Indeed, the right inverse of a function exists if the function is surjective, but we can make the function surjective by restricting its codomain to its image. Moreover, an advantage of this formulation is that it is not required to compute the decoders explicitly because we only need the compositions $Q^\HF \circ \mathcal D^\HF$ and $Q^\LF \circ \mathcal D^\LF$, which correspond to the identity function by equations \eqref{eq:condition_particular_AE}, and therefore we do not need to train the autoencoders. In this case, the modified low-fidelity model simplifies to
\begin{equation}
\mathcal Q_\AE^\LF(\bm \xi^\HF) = (\mathcal S^\LF)^{-1}(\mathcal S^\HF(Q^\HF(\bm \xi^\HF); \sigma^\HF); \sigma^\LF).
\end{equation}
The only limitation of this approach is that the new low-fidelity model $\mathcal Q_\AE^\LF$ depends on the high-fidelity model $Q^\HF$, which must therefore be replaced by a cheaper surrogate model based on fully connected neural networks, as already done for the other methods and described in \cref{rem:surrogate_AS,rem:surrogate_AE}. On the other hand, when the quantity of interest is scalar, the fact that the encoder is the model itself implies that the shared space is one-dimensional, which in turn yields that the normalizing flows $\mathcal S^\HF(\cdot; \sigma^\HF)$ and $\mathcal S^\LF(\cdot; \sigma^\LF)$ are one-dimensional mappings.

\subsection{Computational complexity analysis} \label{sec:analysis_cost}

In this section we analyze the computational budget $\mathfrak C$ that we need to spend in order to perform our pipelines. It is clear that this additional computational cost is not required for the standard multifidelity Monte Carlo estimator. Hence, one can argue that this budget might be better invested in high-fidelity samples rather than in methodologies for increasing the correlation between the models. In the next proposition we study when it is worth spending part of the computational budget in building our modified estimators. The result is dependent on the ratio $\eta = \mathfrak C / \mathcal B$ between the cost for training the networks in the pipelines and the total computational budget, and on the improvement in term of correlation between the models.

\begin{proposition}
Let $\rho$ be the initial correlation between the high-fidelity and low-fidelity models, and let $\rho_\AS$ and $\rho_\AE$ be new correlations after performing the pipelines based on active subspace and autoencoder, respectively. Assume that
\begin{equation}
\abs{\rho_\A} > \abs{\rho} > \frac{4w}{(1+w)^2},
\end{equation}
and
\begin{equation}
\frac{\mathfrak C}{\mathcal B} = \eta < 1 - \frac{\sqrt{1 - \rho_\A^2} + \sqrt{w \rho_\A^2}}{\sqrt{1 - \rho^2} + \sqrt{w \rho^2}},
\end{equation}
where $\mathfrak C$ is the cost for training the networks, $\mathcal B$ is the total computational budget, $\A$ stands for both $\AS$ and $\AE$, and $w = \mathcal C^\LF / \mathcal C^\HF$ is the cost ratio between the two fidelities. Then, it holds
\begin{equation}
\Var \left[ \widehat Q_{\mathcal B}^{\MFMC, \A} \right] < \Var \left[ \widehat Q_{\mathcal B}^{\MFMC} \right],
\end{equation}
where both the estimators are computed assuming a total computational budget $\mathcal B$ and solving the optimal allocation problem.
\end{proposition}
\begin{proof}
Using equation \eqref{eq:variance_MFMC} we have
\begin{equation} \label{eq:proof_variance_MFMC}
\Var \left[ \widehat Q_{\mathcal B}^{\MFMC} \right] = \Var \left[ \widehat Q_{\mathcal B}^{\MC} \right] \left( \sqrt{1 - \rho^2} + \sqrt{w \rho^2} \right) = \frac{\Var \left[ Q^\HF(\bm \xi^\HF) \right]}{\mathcal B} \left( \sqrt{1 - \rho^2} + \sqrt{w \rho^2} \right).
\end{equation}
Moreover, since the computational budget $\mathfrak C = \eta \mathcal B$ is spent for building the modified estimators, we then employ equation \eqref{eq:variance_MFMC} with budget $\mathcal B - \mathfrak C$, and obtain
\begin{equation} \label{eq:proof_variance_MFMC_A}
\begin{aligned}
\Var \left[ \widehat Q_{\mathcal B}^{\MFMC,\A} \right] &= \Var \left[ \widehat Q_{\mathcal B - \mathfrak C}^{\MC} \right] \left( \sqrt{1 - \rho_\A^2} + \sqrt{w \rho_\A^2} \right) \\
&= \frac{\Var \left[ Q^\HF(\bm \xi^\HF) \right]}{\mathcal B - \mathfrak C} \left( \sqrt{1 - \rho_\A^2} + \sqrt{w \rho_\A^2} \right) \\
&= \frac{\Var \left[ Q^\HF(\bm \xi^\HF) \right]}{\mathcal B(1 - \eta)} \left( \sqrt{1 - \rho_\A^2} + \sqrt{w \rho_\A^2} \right).
\end{aligned}
\end{equation}
Finally, combining equations \eqref{eq:proof_variance_MFMC} and \eqref{eq:proof_variance_MFMC_A} gives the desired result.
\end{proof}

The previous result shows the condition under which we get a benefit in using our approaches. In particular, as long as the correlation increases significantly and the cost for building the modified estimators is sufficiently small compared to the total computational budget, it is better to apply our methodologies rather than using standard multifidelity Monte Carlo. This is true especially for all computationally expensive models that appear in concrete applications, such as the cardiovascular simulations in \cref{sec:cardio}.

\subsection{A theoretical example} \label{sec:theoretical_example}

\begin{figure}[t]
\centering
\begin{tabular}{ccccc}
\includegraphics{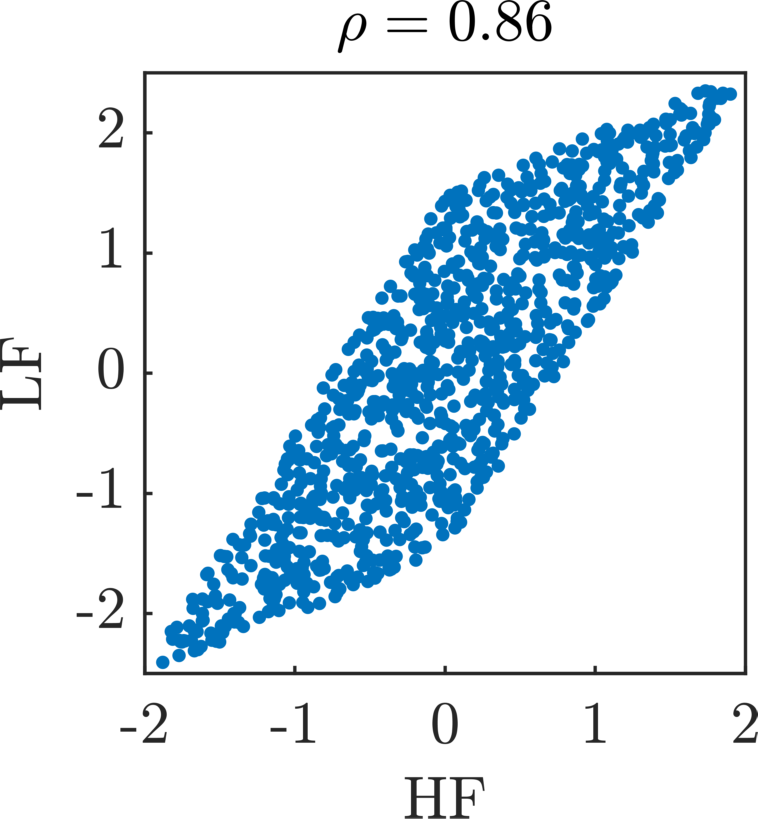} & &
\includegraphics{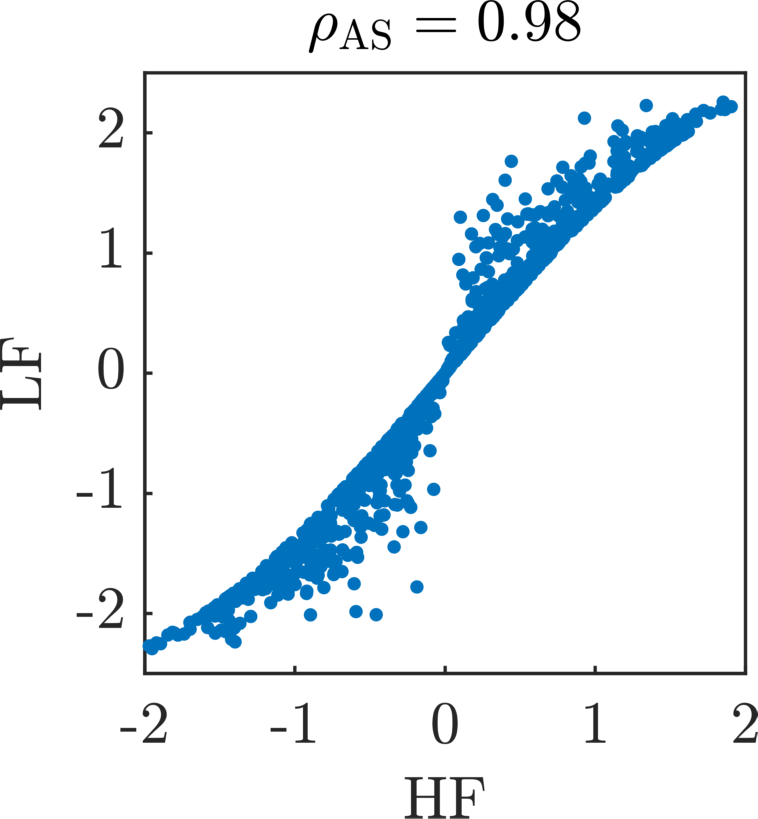} & &
\includegraphics{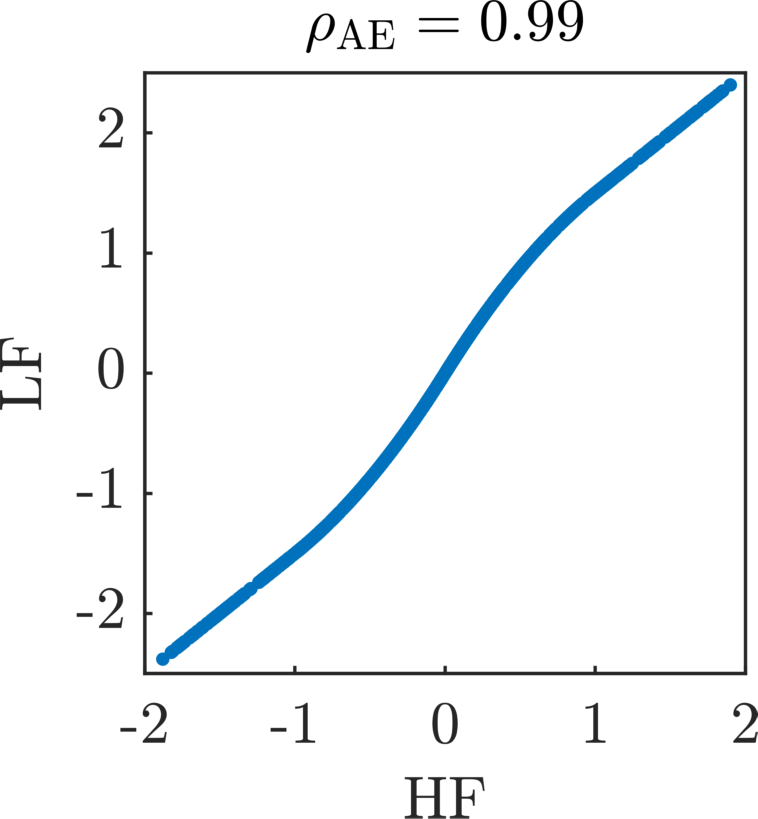}  \\
\includegraphics{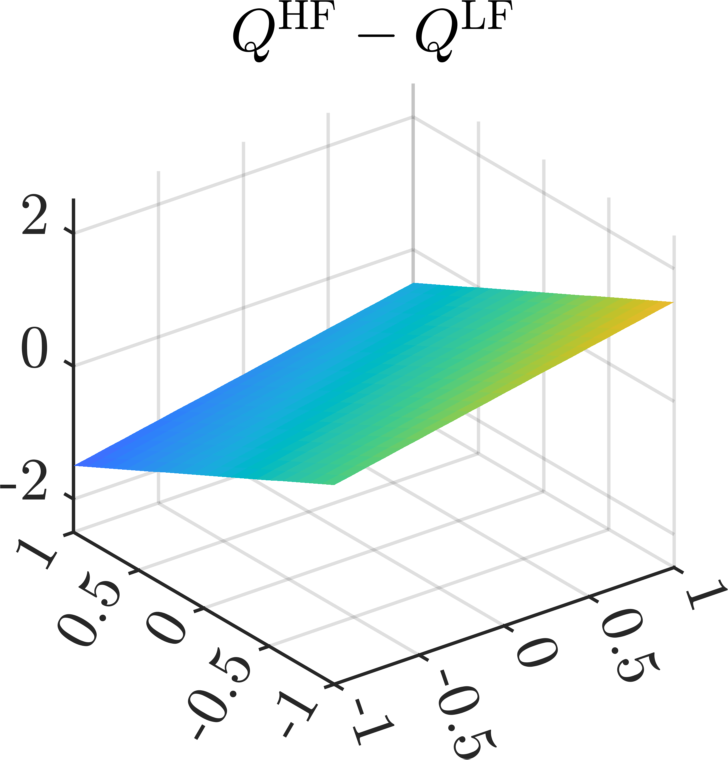} & &
\includegraphics{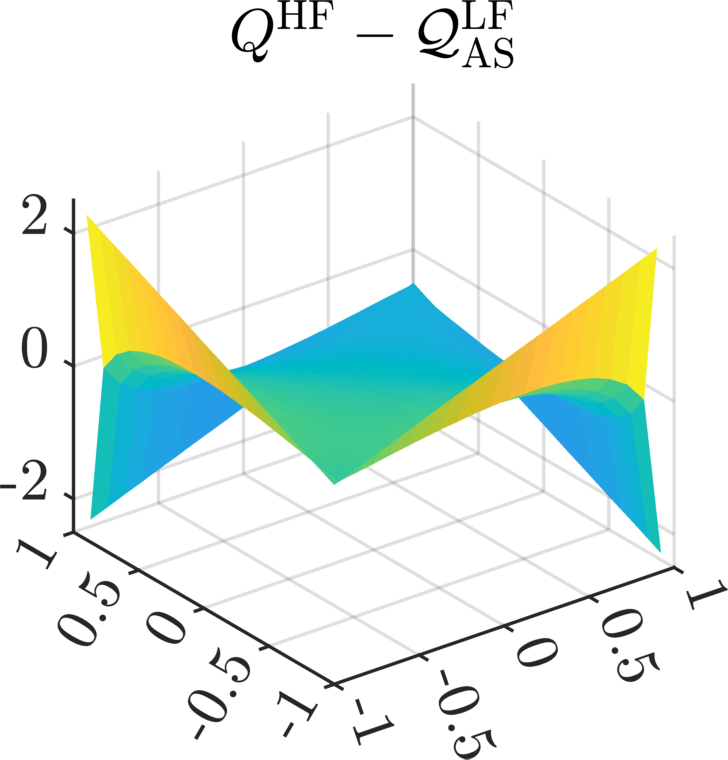} & &
\includegraphics{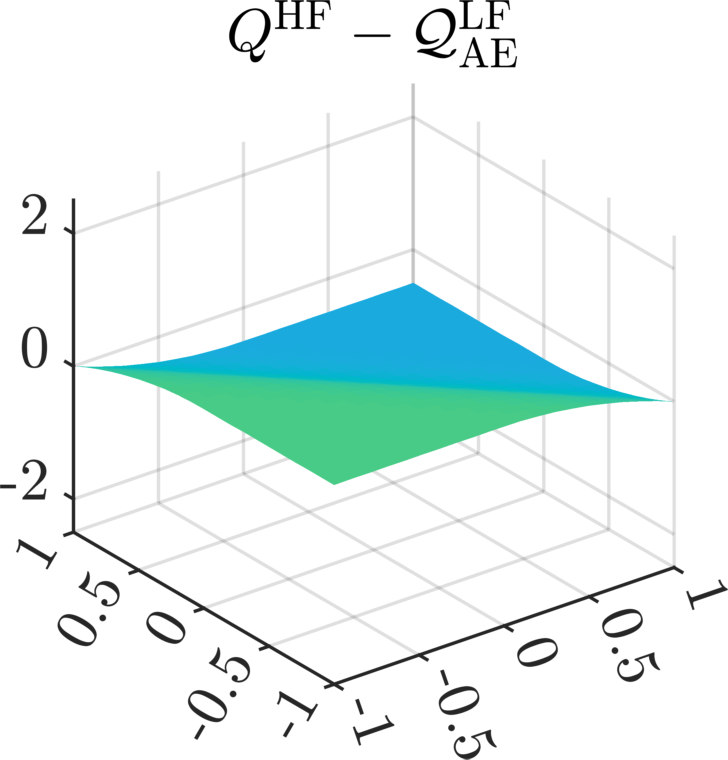} 
\end{tabular}
\caption{Correlation (top) and difference (bottom) between HF model and original LF model (left), LF models given by the methods based on active subspaces (center) and autoencoders (right).}
\label{fig:analytic_example_correlation}
\end{figure}

\begin{figure}[t]
\centering
\includegraphics{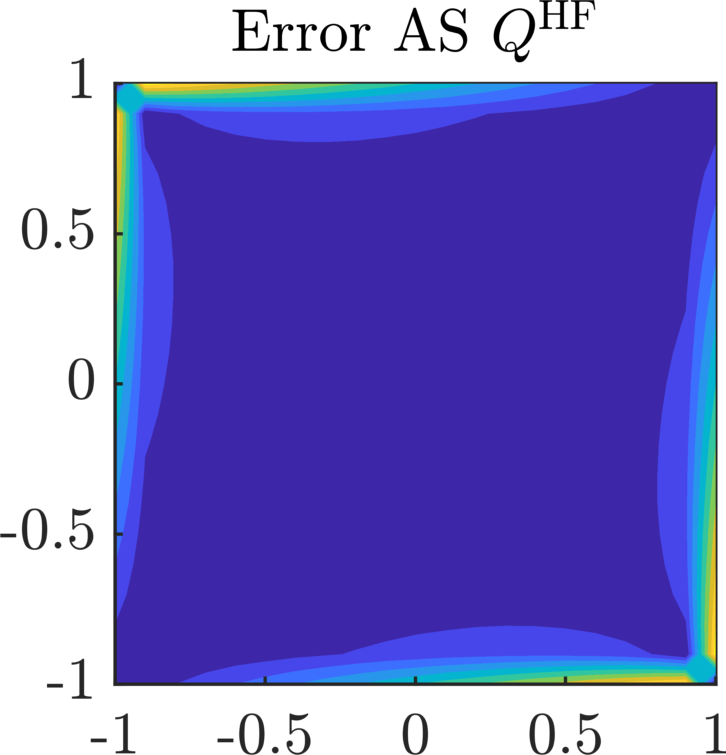}
\includegraphics{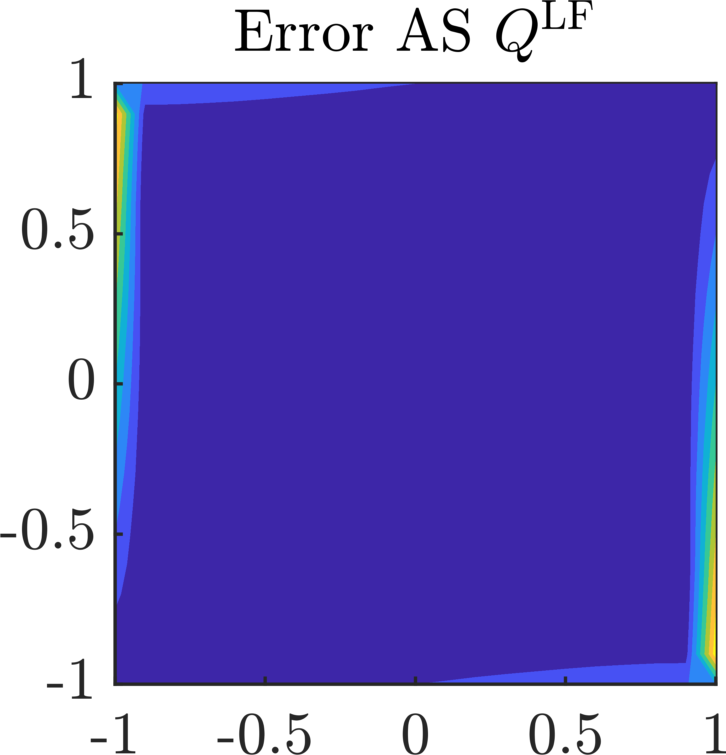}
\includegraphics{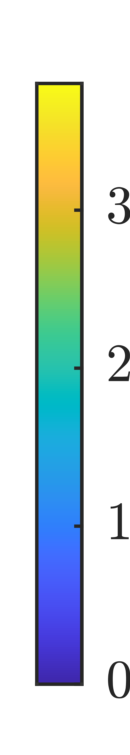}
\caption{Truncation error introduced reducing the dimensionality of the models through active subspaces and normalizing flow, for the theoretical example. The error is computed as $\abs{Q^\mathrm{*F}(x,y) - Q^\mathrm{*F}(\mathcal T^{-1}(W_A^\mathrm{*F} (W_A^\mathrm{*F})^\top \mathcal T(x, y)))}$, where $\mathrm{*F}$ stands for both $\HF$ and $\LF$.}
\label{fig:analytic_example_errorAS}
\end{figure}

\begin{figure}[t]
\centering
\includegraphics{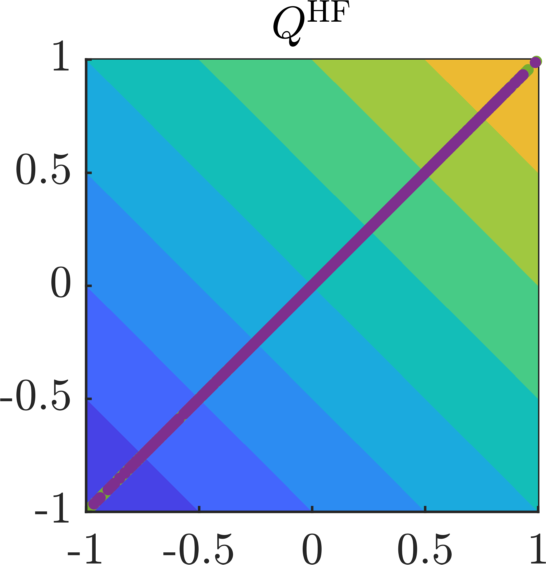}
\includegraphics{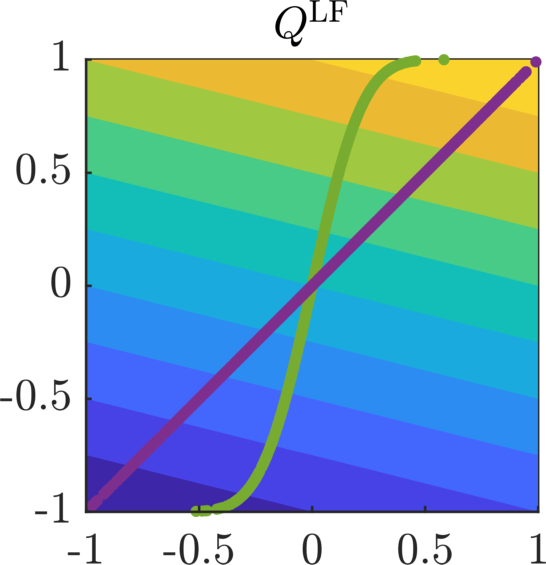}
\includegraphics{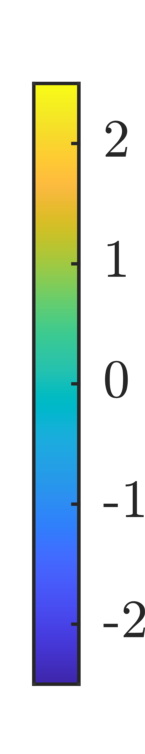} $\qquad$
\includegraphics{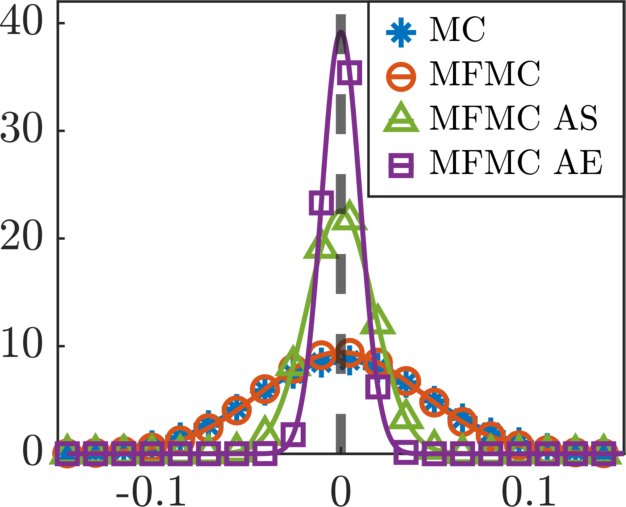}
\caption{Left: contour plot of $Q^\HF$ and $Q^\LF$ with the corresponding dominant directions determined by the methods based on active subspaces (in green) and autoencoders (in purple), for the theoretical example. Right: comparison between our methods (MFMC AS and MFMC AE) with standard (multifidelity) Monte Carlo (MC and MFMC), for the theoretical example.}
\label{fig:analytic_example_reduced}
\end{figure}

The goal of this section is to give a better understanding of the methodologies introduced in the previous sections, by considering a simple example where every computation can be performed analytically. Let the high-fidelity and low-fidelity models $Q^\HF, Q^\LF \colon \R^2 \to \R$ be the functions
\begin{equation}
Q^\HF(x,y) = x + y \qquad \text{and} \qquad Q^\LF(x,y) = \frac{x}2 + 2y,
\end{equation}
and let the distributions of the input values be $\mu^\HF = \mu^\LF = \mu = \mathcal U([-1,1]^2)$. First, notice that the two models have zero mean and consequently their Pearson correlation coefficient is
\begin{equation}
\rho = \frac{\E \left[ \left( X + Y \right) \left( \frac{X}2 + 2Y \right) \right]}{\sqrt{\E[(X+Y)^2] \E \left [ \left(\frac{X}2+2Y \right)^2 \right]}} = \frac{5}{\sqrt{34}} \simeq 0.86.
\end{equation}
We now apply our approaches to modify the low-fidelity model and thus increase the correlation coefficient. Let us first consider the method presented in \cref{sec:MFMC_AS}. The normalizing flow, which transforms the uniform distribution $\mu$ into a standard two-dimensional Gaussian $\mathcal N(0, I_2)$, and its inverse are given by
\begin{equation}
\mathcal T(x,y) = \begin{pmatrix} \sqrt 2 \erf^{-1}(x) \\ \sqrt 2 \erf^{-1}(y) \end{pmatrix} \qquad \text{and} \qquad \mathcal T^{-1}(x,y) = \begin{pmatrix} \erf \left(\frac{x}{\sqrt 2} \right) \\ \erf \left(\frac{y}{\sqrt 2} \right) \end{pmatrix},
\end{equation}
which is equal for both the high-fidelity  and low-fidelity models. We then obtain the modified models $\widetilde Q^\HF, \widetilde Q^\LF \colon \R^2 \to \R$
\begin{equation}
\begin{aligned}
\widetilde Q^\HF(x,y) &= Q^\HF(\mathcal T^{-1}(x,y)) = \erf \left(\frac{x}{\sqrt 2} \right) + \erf \left(\frac{y}{\sqrt 2} \right), \\
\widetilde Q^\LF(x,y) &= Q^\LF(\mathcal T^{-1}(x,y)) = \frac12 \erf \left(\frac{x}{\sqrt 2} \right) + 2 \erf \left(\frac{y}{\sqrt 2} \right),
\end{aligned}
\end{equation}
whose input distribution is the standard Gaussian, and which give the matrices
\begin{equation}
C^\HF = \E \left[ \nabla \widetilde Q^\HF(X,Y) \nabla \widetilde Q^\HF (X,Y) ^\top \right] = \frac2\pi \E \left[ \begin{pmatrix} e^{-X^2} & e^{- \frac{X^2 + Y^2}2} \\ e^{- \frac{X^2 + Y^2}2} & e^{- Y^2} \end{pmatrix} \right] = \begin{pmatrix} \frac{2}{\pi \sqrt 3} & \frac1\pi \\ \frac1\pi & \frac2{\pi \sqrt 3} \end{pmatrix},
\end{equation}
and
\begin{equation}
C^\LF = \E \left[ \nabla \widetilde Q^\LF(X,Y) \nabla \widetilde Q^\LF (X,Y) ^\top \right] = \frac2\pi \E \left[ \begin{pmatrix} \frac14 e^{-X^2} & e^{- \frac{X^2 + Y^2}2} \\ e^{- \frac{X^2 + Y^2}2} & 4 e^{- Y^2} \end{pmatrix} \right] = \begin{pmatrix} \frac{1}{2 \pi \sqrt 3} & \frac1\pi \\ \frac1\pi & \frac8{\pi \sqrt 3} \end{pmatrix}.
\end{equation}
The corresponding one-dimensional active subspaces are
\begin{equation}
W_A^\HF = \frac1{\sqrt2} \begin{pmatrix} 1 \\ 1\end{pmatrix} \qquad \text{and} \qquad W_A^\LF = \frac1{\sqrt{273 + 15\sqrt{273}}} \begin{pmatrix} 2 \sqrt6 \\ \frac{15 + \sqrt{273}}{\sqrt2} \end{pmatrix},
\end{equation}
and are used in the definition of the new low-fidelity model
\begin{equation}
\mathcal Q^\LF_\AS(x, y) = Q^\LF(\mathcal T^{-1}(W_A^\LF (W_A^\HF)^\top \mathcal T(x, y))),
\end{equation}
which yields an improved correlation 
\begin{equation}
\rho_\AS = \frac{\Cov (Q^\HF(X,Y), \mathcal Q_\AS^\LF(X,Y))}{\sqrt{\Var[Q^\HF(X,Y)] \Var[\mathcal Q_\AS^\LF(X,Y)]}} \simeq 0.98 > \rho.
\end{equation}
We remark that the correlation coefficient is computed numerically employing $10^8$ samples from the distribution $\mu$. Let us now focus on the method presented in \cref{sec:MFMC_AE}. Consider the following choice for the autoencoders
\begin{equation}
\begin{aligned}
&\begin{cases}
\mathcal E^\HF \colon [-1,1]^2 \to [-2,2], & \mathcal E^\HF(x,y) = x + y, \\
\mathcal D^\HF \colon [-2,2] \to [-1,1]^2, & \mathcal D^\HF(z) = \begin{pmatrix} \frac{z}2 \\ \frac{z}2 \end{pmatrix},
\end{cases} \\
&\begin{cases}
\mathcal E^\LF \colon [-1,1]^2 \to \left[-\frac52,\frac52\right], & \mathcal E^\LF(x,y) = \frac{x}2 + 2y, \\
\mathcal D^\LF \colon \left[-\frac52,\frac52\right] \to [-1,1]^2, & \mathcal D^\LF(z) = \begin{pmatrix} \frac25 z \\ \frac25 z \end{pmatrix},
\end{cases}
\end{aligned}
\end{equation}
and notice that the original models can be reconstructed exactly, i.e., 
\begin{equation}
Q^\HF(x,y) = Q^\HF(\mathcal D^\HF(\mathcal E^\HF(x,y))) \qquad \text{and} \qquad Q^\LF(x,y) = Q^\LF(\mathcal D^\LF(\mathcal E^\LF(x,y))).
\end{equation}
We remark that the autoencoders which give an exact representation of the functions are not unique, for example we could rescale the encoders by a constant $c>0$ and compute the decoders accordingly, without affecting the final correlation. We now have to find a normalizing flow from the latent space of each model to a standard one-dimensional Gaussian $\mathcal N(0,1)$. From the encoders $\mathcal E^\HF$ and $\mathcal E^\LF$, we deduce that the distributions of the latent spaces of the high-fidelity and low-fidelity models are
\begin{equation}
\nu^\HF = \mathcal Tri(-2,0,+2) \qquad \text{and} \qquad \nu^\LF = \mathcal Trap \left( -\frac52, -\frac32, +\frac32, +\frac52 \right),
\end{equation}
where $\mathcal Tri$ and $\mathcal Trap$ stand for triangular and trapezoidal distribution, respectively. Hence, following \cite{Gra13}, the normalizing flows are given by
\begin{equation}
\mathcal S^\HF(z) = \sqrt2 \erf^{-1}(U^\HF(z)) \qquad \text{and} \qquad \mathcal S^\LF(z) = \sqrt2 \erf^{-1}(U^\LF(z)),
\end{equation}
where
\begin{equation}
\begin{aligned}
U^\HF(z) &= \begin{cases}
\frac14 (2+z)^2 - 1, & -2 \le z \le 0, \\
1 - \frac14 (2-z)^2, & 0 \le z \le 2,
\end{cases} \\
U^\LF(z) &= \begin{cases}
\frac14 \left( \frac52 + z \right)^2 - 1, & -\frac52 \le z \le -\frac32, \\
\frac{z}2, & -\frac32 \le z \le \frac32, \\
1 - \frac14 \left( \frac52 - z \right)^2, & \frac32 \le z \le \frac52,
\end{cases}
\end{aligned}
\end{equation}
These transformations are used in the definition of the new low-fidelity model
\begin{equation}
\mathcal Q_\AE^\LF(x,y) = Q^\LF(\mathcal D^\LF((\mathcal S^\LF)^{-1}(\mathcal S^\HF(\mathcal E^\HF(x,y))))),
\end{equation}
which yields an improved correlation 
\begin{equation}
\rho_\AE = \frac{\Cov (Q^\HF(X,Y), \mathcal Q_\AE^\LF(X,Y))}{\sqrt{\Var[Q^\HF(X,Y)] \Var[\mathcal Q_\AE^\LF(X,Y)]}} \simeq 0.99 > \rho_\AS > \rho.
\end{equation}
The correlation coefficient is computed numerically employing $10^8$ samples from the distribution $\mu$, in the same way we did for the method based on active subspaces. In \cref{fig:analytic_example_correlation} we plot the correlation between the high-fidelity model and both the original and the new low-fidelity models, together with their difference. We notice that employing a nonlinear transformation results in a larger correlation coefficient, i.e., $\rho_\AE > \rho_\AS$, which cannot be obtained by means of a linear transformation, and a better approximation of the high-fidelity model, since the differnce is closer to zero. These plots also show that active subspaces introduce a small truncation error, which is not produced by the autoencoder, as seen in \cref{fig:analytic_example_errorAS}. Moreover, in \cref{fig:analytic_example_reduced} we plot the dominant subspaces obtained by employing our methodologies. We observe that the two approaches find the same subspace for the high-fidelity model, while they provide different subspaces for the low-fidelity one. The latter is therefore responsible for the better correlation between the models. Finally, still in \cref{fig:analytic_example_reduced} we compare standard Monte Carlo (MC) and multifidelity Monte Carlo (MFMC) with our two approaches (MFMC AS) and (MFMC AE). We assume a ratio between the costs of the models equal to $\mathcal C^\LF = 0.01 \mathcal C^\HF$ to mirror cost differences in realistic applications, and we then set a budget of $100$ HF simulations and $20000$ LF simulations, which is equivalent to the cost of $300$ HF simulations. We observe that both our methods outperform standard techniques, and, in particular, the methodology with the autoencoder achieves a smaller variance with respect to the active subspace due to a larger correlation between the high-fidelity and the reduced low-fidelity models.

\begin{remark}
Notice that equation \eqref{eq:variance_MFMC} still holds true for the estimators in \eqref{eq:estimator_MFMCAS} and \eqref{eq:estimator_MFMCAE}. Therefore, if we manage to increase the correlation between the high-fidelity and low-fidelity models, then we also improve the variance of the resulting estimators. For the active subspace technique, and in general for linear approaches, it is reasonable to assume that if we align the important directions of different models, then their correlation should be larger along those directions than in the original space. The intuition for this approach is provided in previous literature~\cite{GEG18,ZGE23}; specifically, in \cite[Proposition 4.4]{ZGE23}, it is shown how the re-arrangement of the variables leads to an increased correlation in the linear case. Beyond the intuition or the quantitative analysis presented in~\cite{ZGE23} under simplifying assumptions, we also note this idea has been adopted successfully on non-trivial aerospace applications with high-dimensional inputs, see, e.g.,~\cite{GeE18,GEGJ19}. On the other hand, even if we do not have any theoretical guarantee that nonlinear lower-dimensional manifolds can increase the correlation, we expect them to behave similarly if mapped appropriately. In this work, we consider autoencoders as a nonlinear extension to linear dimensionality reduction, and the following numerical experiments highlight the possibility to improve the performance of multifidelity Monte Carlo estimators whenever a nonlinear manifold provides a more parsimonious representation than a linear subspace for the input-to-output map in at least one of the models. Moreover, we can always verify whether the new correlation, which can be estimated through a pilot sample as demonstrated in the paper, is larger than the original correlation. If this does not hold, then we can employ standard multifidelity Monte Carlo estimators, so that we can guarantee no reduction in performance apart from the negligible cost increase (compared to the models' evaluations) of the dimension reduction and normalizing flows steps. We finally note that one could train the autoencoders for both the high-fidelity and low-fidelity models simultaneously, and including a term in the loss function that maximizes the resulting correlation.
\end{remark}

\section{Numerical experiments} \label{sec:numerical_experiments}

In this section we demonstrate the advantages of our approaches through a series of test cases. We first consider analytic functions which allow us to explore the properties of the methods, and then focus on a reaction-diffusion equation. Finally, we consider cardiovascular simulations as an example of a computationally expensive model with concrete applications and for which only a small amount of data can be available. We note that the dashed lines representing the ``true'' mean in the following plots is given by the average of the Monte Carlo estimator.

\begin{remark} \label{rem:hyperparameters}
For all the neural networks appearing in the pipelines, after normalizing the input and output values in the interval $[-1,1]$, we perform a hyperparameter tuning for the number of layers, number of neurons per layer, learning rate, and exponential scheduler step. In particular, we tune the hyperparamters which appear in the autoencoder, the surrogate models, and the normalizing flows. The hyperparameters are optimized sequentially employing the Optuna optimization framework \cite{ASY19} monitoring the validation loss (20\% of the dataset). In particular, we initially find the best parameters for the surrogate models, if necessary, and then use these parameters in the training process for the autoencoder. Then, once we have selected the best parameters for the autoencoder, we employ these values to compute the latent space from which we learn the normalizing flows and their corresponding hyperparameters. In the following numerical experiments, we constrain the number of layers in $\{ 1, \dots, 4 \}$, the number of neurons per layer in $\{ 1, \dots, 16 \}$, the learning rate in $[ 10^{-4}, 10^{-2} ]$, and the exponential scheduler step in $[ 0.999, 0.9999 ]$. Moreover, we train all the neural networks for 5000 epochs with the Adam optimizer \cite{KiB17}, and we perform 100 independent repetitions of the entire procedure in order to illustrate its overall variability. The major computational cost for obtaining the networks is given by the hyperparameter tuning, which is however done as a preliminary step, before performing the multifidelity uncertainty propagation pipeline. We also notice that, as highlighted in \cref{sec:analysis_cost}, the cost for training all the networks in the pipelines is negligible with respect to the cost of high-fidelity and low-fidelity simulations, in particular for computationally expensive models, and therefore we do not include this cost in the comparison between our approaches and standard (multifidelity) Monte Carlo estimators.
\end{remark}

\subsection{Analytic functions}

\begin{figure}[t]
\centering
\begin{tabular}{cccc}
Analytic NF & Analytic NF & Spline NF & Spline NF \\
Actual model & Surrogate model & Actual model & Surrogate model \\
\includegraphics{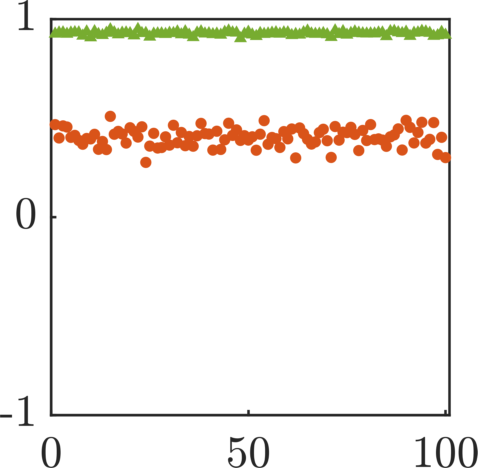} &
\includegraphics{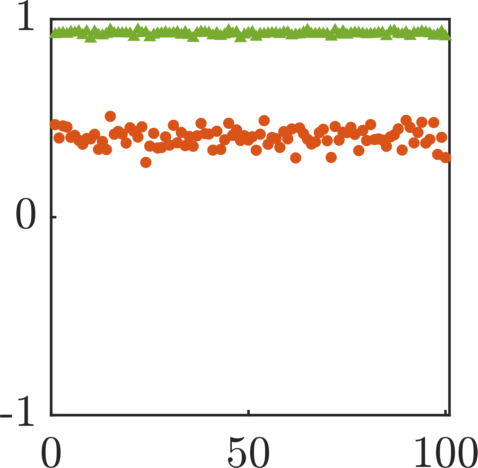} &
\includegraphics{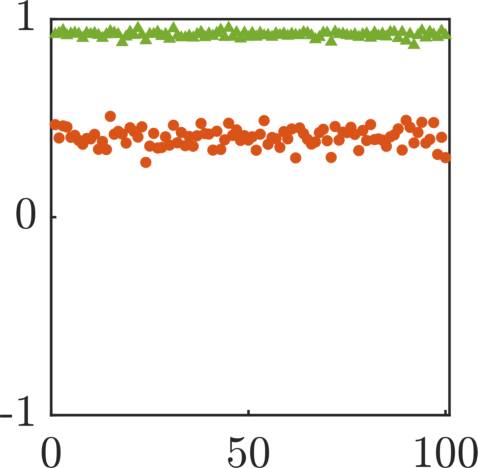} &
\includegraphics{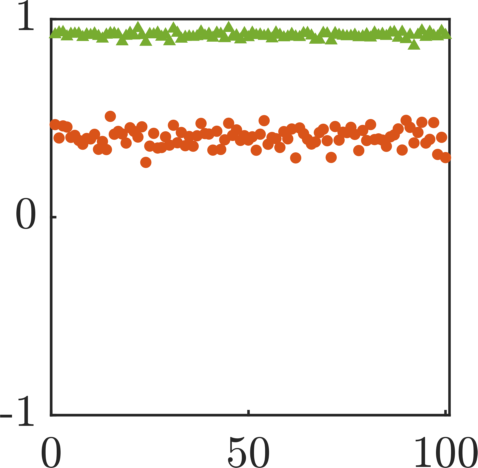} \\
\includegraphics{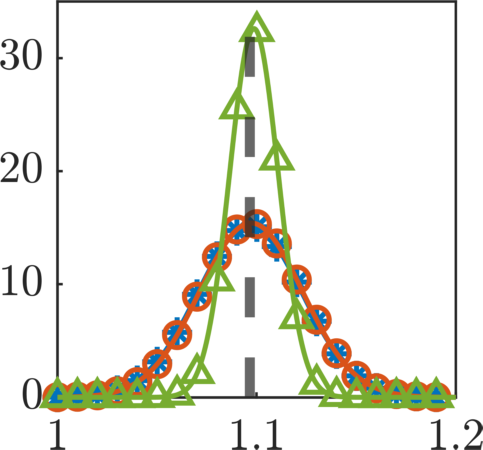} &
\includegraphics{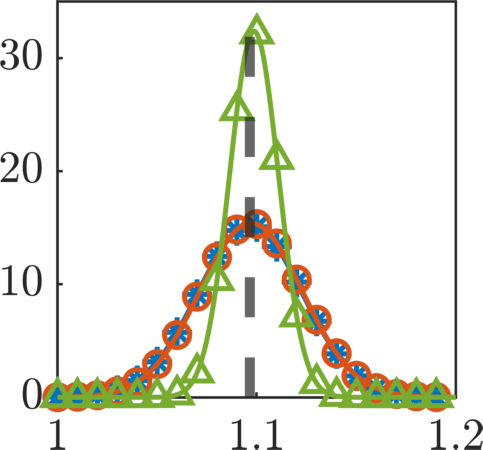} &
\includegraphics{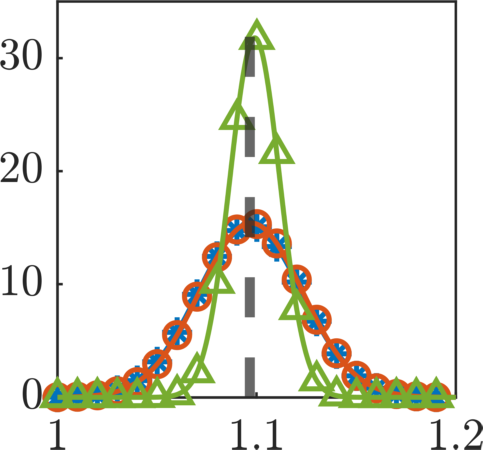} &
\includegraphics{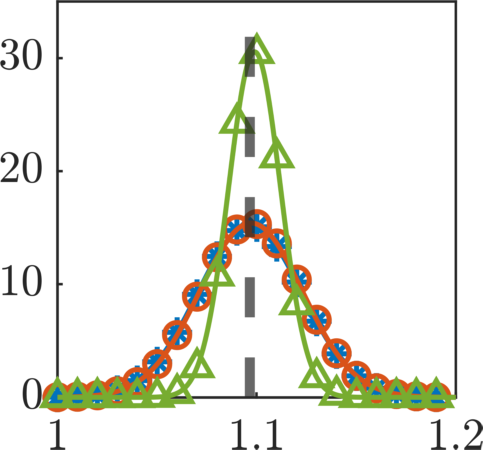}
\end{tabular}
\includegraphics[scale=0.2]{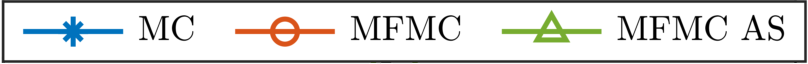}
\caption{Comparison between our method based on active subspaces (MFMC AS) with standard (multifidelity) Monte Carlo (MC and MFMC), for the case of analytic functions. The normalizing flow can be either exact (Analytic NF) or estimated using splines (Spline NF), and the gradient to compute the active subspace can be either obtained with the analytic gradient (Actual model) or through a surrogate model given by a neural network (Surrogate model). Top: Pearson correlation coefficients for 100 different repetitions. Bottom: approximated distributions of the estimators using 100 samples}
\label{fig:analytic_functions_AS}
\end{figure}

\begin{figure}[t]
\centering
\begin{tabular}{cccc}
\includegraphics{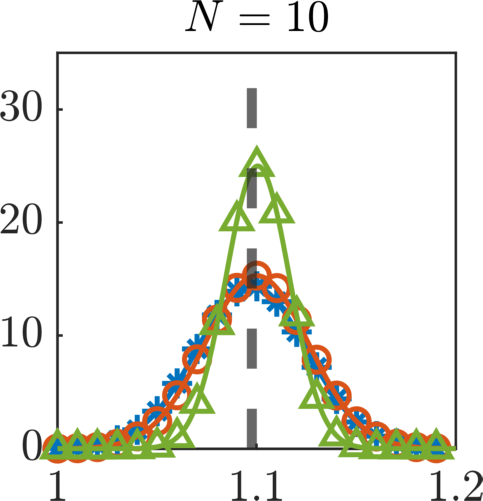} &
\includegraphics{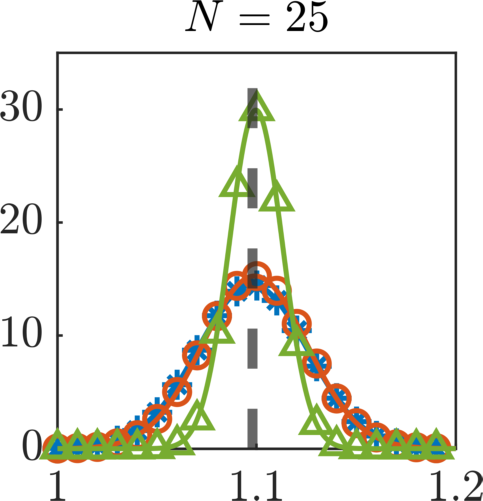} &
\includegraphics{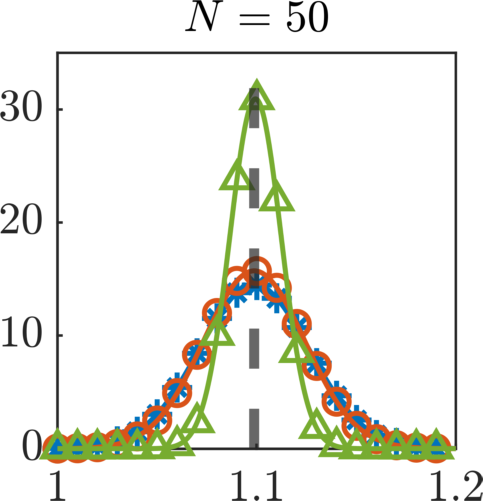} &
\includegraphics{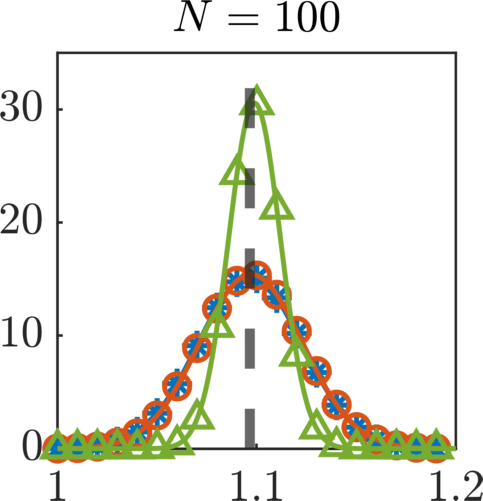}
\end{tabular}
\includegraphics[scale=0.2]{figures/legend_AS_horizontal}
\caption{Comparison between our method based on active subspaces (MFMC AS) with standard (multifidelity) Monte Carlo (MC and MFMC), varying the number of data points, for the case of analytic functions. The normalizing flow is estimated using splines, and the gradient to compute the active subspace is obtained through a surrogate model given by a neural network.}
\label{fig:analytic_functions_AS_N}
\end{figure}

\begin{figure}[t]
\centering
\begin{tabular}{cccc}
Model as encoder & Model as encoder & Classic AE & Classic AE \\
Actual model & Surrogate model & Actual model & Surrogate model \\
\includegraphics{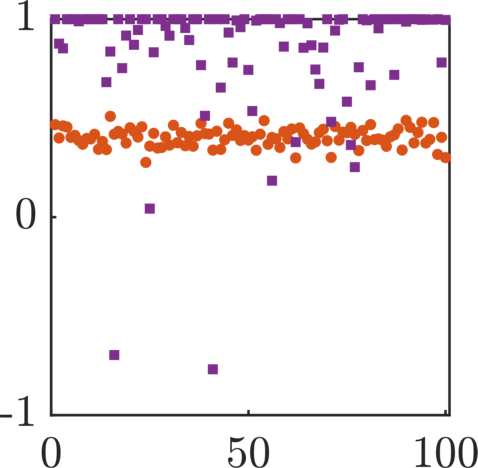} &
\includegraphics{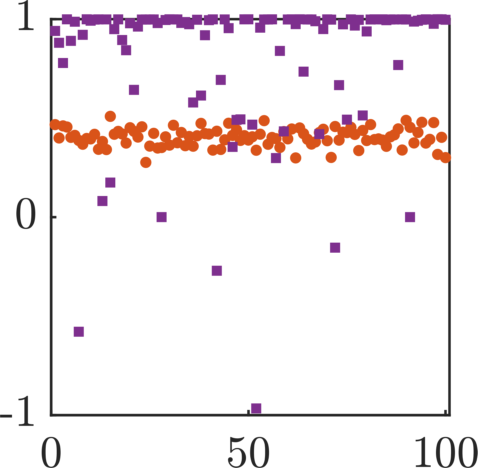} &
\includegraphics{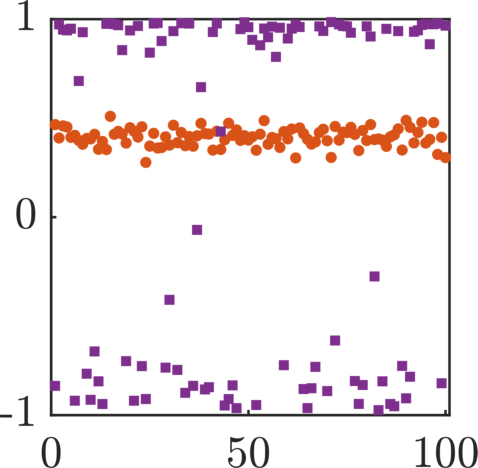} &
\includegraphics{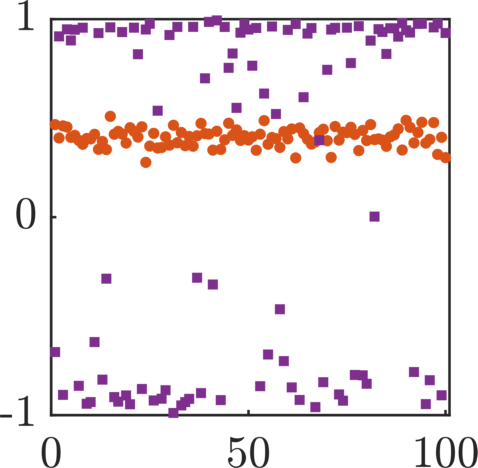} \\
\includegraphics{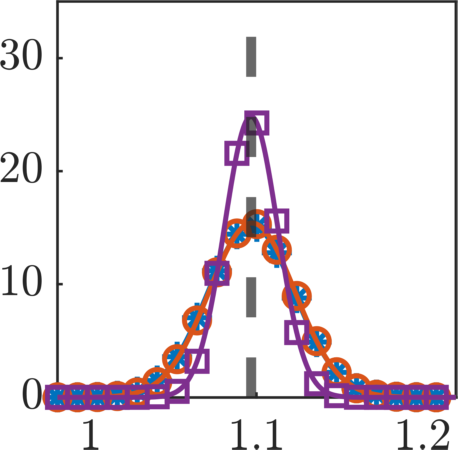} &
\includegraphics{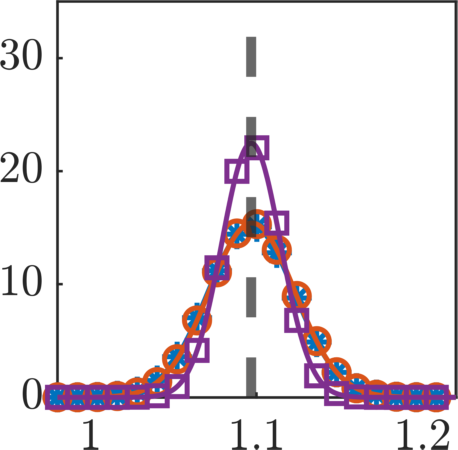} &
\includegraphics{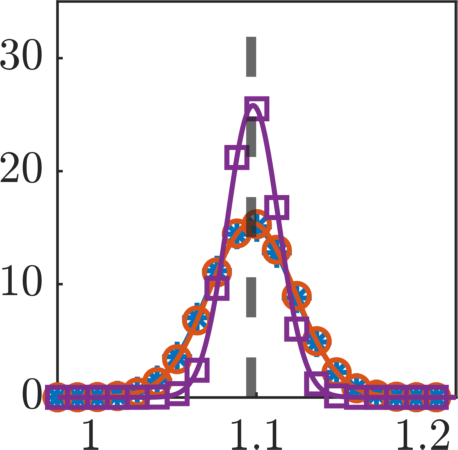} &
\includegraphics{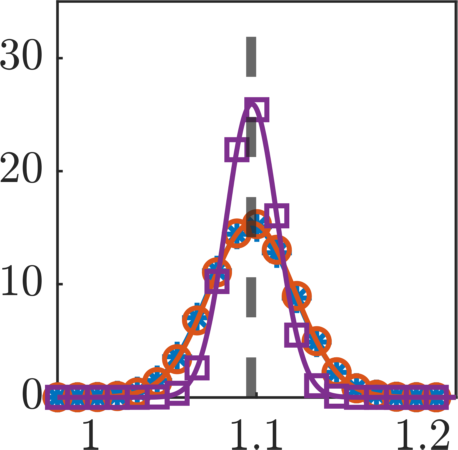}
\end{tabular}
\includegraphics[scale=0.2]{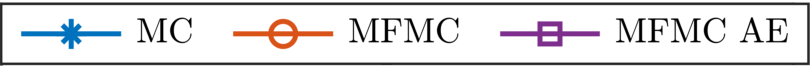}
\caption{Comparison between our method based on autoencoders (MFMC AE) with standard (multifidelity) Monte Carlo (MC and MFMC), for the case of analytic functions. The encoder can be either fixed (Model as encoder) or computed (Classic AE), and for the training of the autoencoder we can use either the function (Actual model) or a surrogate model given by a neural network (Surrogate model). Top: Pearson correlation coefficients for 100 different repetitions. Bottom: approximated distributions of the estimators using 100 samples.}
\label{fig:analytic_functions_AE}
\end{figure}

\begin{figure}[t]
\centering
\begin{tabular}{cccc}
\includegraphics{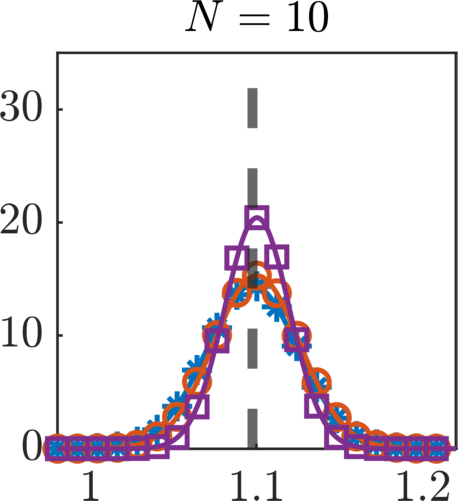} &
\includegraphics{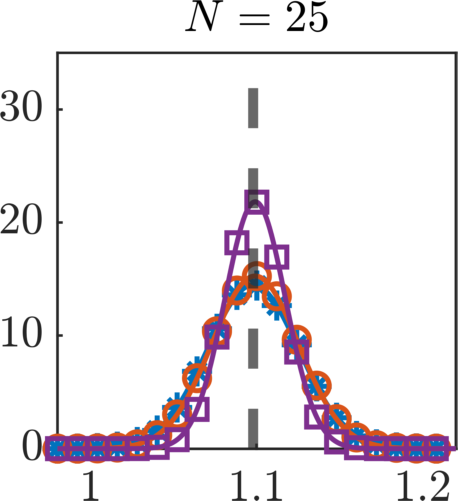} &
\includegraphics{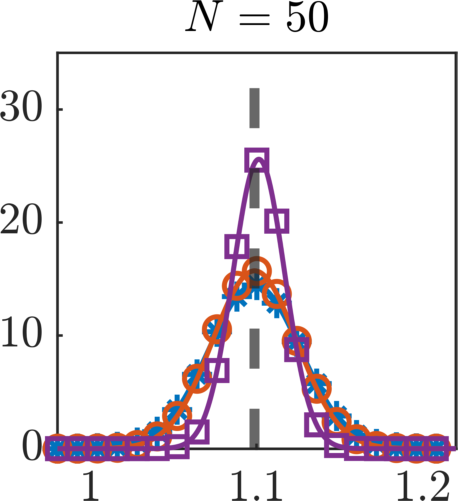} &
\includegraphics{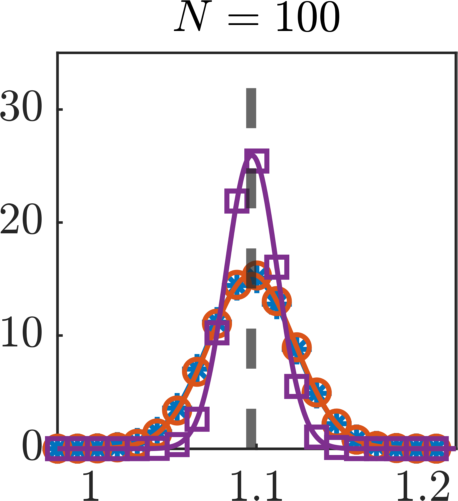}
\end{tabular}
\includegraphics[scale=0.2]{figures/legend_AE_horizontal}
\caption{Comparison between our method based on autoencoders (MFMC AE) with standard (multifidelity) Monte Carlo (MC and MFMC), varying the number of data points, for the case of analytic functions. The autoencoder is trained using a surrogate model given by a neural network}
\label{fig:analytic_functions_AE_N}
\end{figure}

\begin{figure}[t]
\centering
\begin{tabular}{ccccc}
\includegraphics{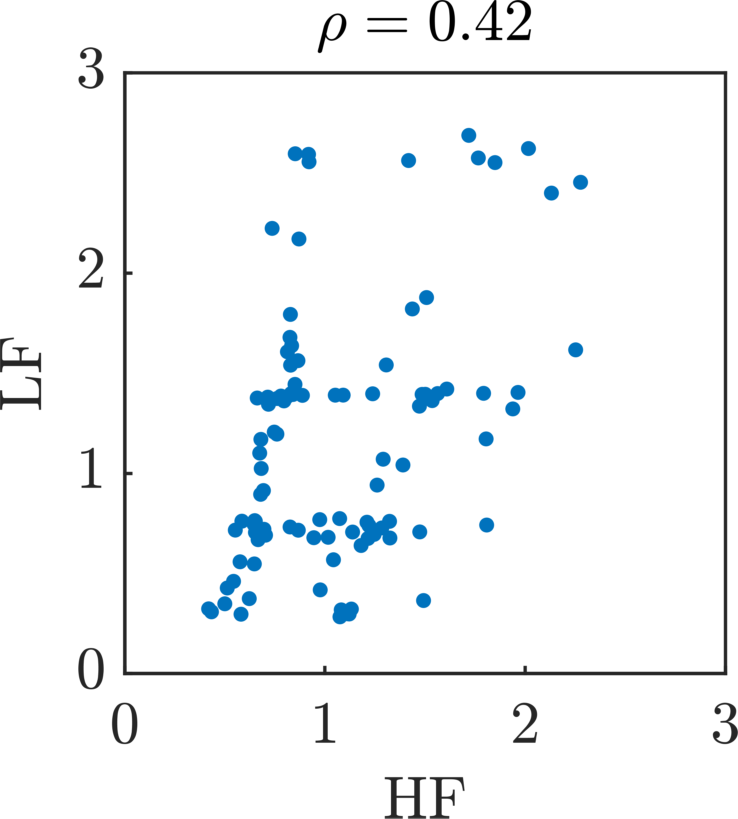} & $\quad$ &
\includegraphics{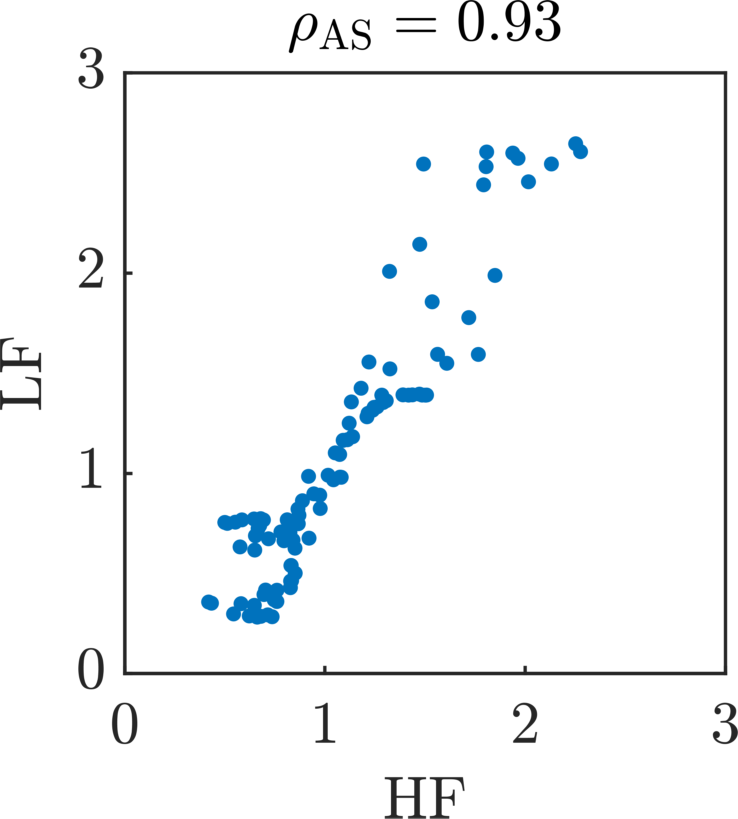} & $\quad$ &
\includegraphics{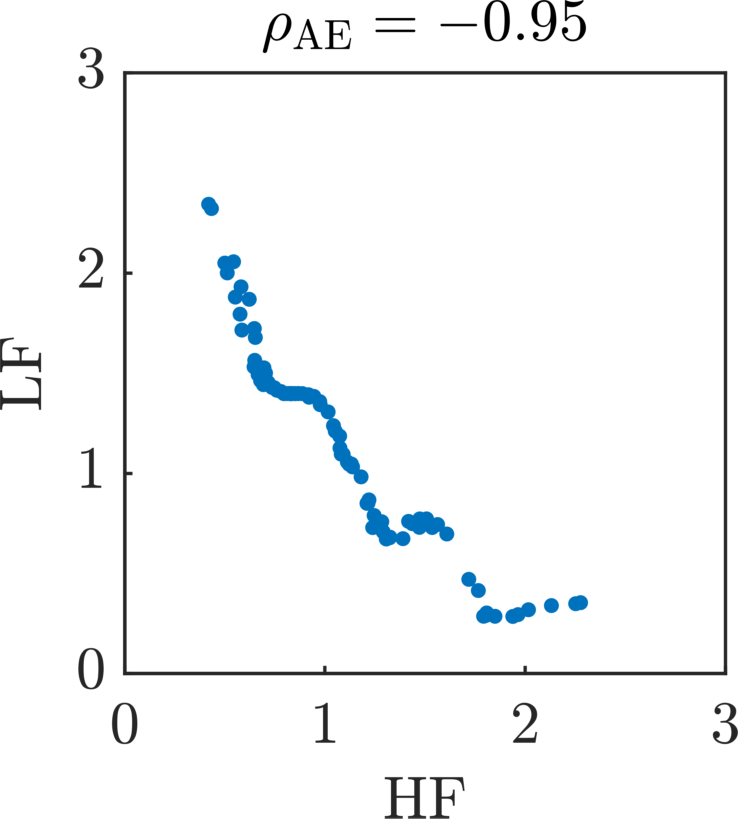}
\end{tabular}
\caption{Correlation between HF model and original LF model (left), LF models given by the methods based on active subspaces (center) and autoencoders (right), for the case of analytic functions.}
\label{fig:analytic_functions_corr}
\end{figure}

Inspired by \cite{GEG18}, we consider the following functions as high-fidelity and low-fidelity models
\begin{equation}
Q^\HF(x, y) = e^{0.7x + 0.3y} + 0.15\sin(2\pi x) \quad \text{and} \quad Q^\LF(x, y) = e^{0.01x + 0.99y} + 0.15\sin(3\pi y),
\end{equation}
with input distributions $\mu^\HF = \mu^\LF = \mu = \mathcal U([-1,1]^2)$, and we aim to estimate
\begin{equation}
\E[Q^\HF(x,y)] = \frac{25}{21} \left( e^{7/10} - e^{-7/10} \right) \left( e^{3/10} - e^{-3/10} \right).
\end{equation}
We assume a cost of our low-fidelity model equal to $\mathcal C^\LF = 0.01 \mathcal C^\HF$ ($w = 0.01$) to mirror cost differences in realistic applications. We notice that in this case gradients and normalizing flow needed for the method presented in \cref{sec:MFMC_AS} can be computed analytically, and are given by
\begin{equation} \label{eq:exact_gradient_AS}
\begin{aligned}
\nabla Q^\HF(x,y) &= \begin{pmatrix}
0.7e^{0.7x + 0.3y} + 0.3\pi\cos(2\pi x) \\
0.3\exp^{0.7x + 0.3y}
\end{pmatrix}, \\
\nabla Q^\LF(x,y) &= \begin{pmatrix}
0.01e^{0.01x + 0.99y} \\
0.99\exp^{0.01x + 0.99y} + + 0.45\pi\cos(3\pi y)
\end{pmatrix},
\end{aligned}
\end{equation}
and
\begin{equation} \label{eq:exact_NF_AS}
\mathcal T^\HF(x,y) = \mathcal T^\LF(x,y) = \mathcal T(x,y) = \begin{pmatrix}
\sqrt2 \erf^{-1}(x) \\ \sqrt2 \erf^{-1}(y)
\end{pmatrix}.
\end{equation}
In this section we study the properties of our methodologies and see how they perform on simple examples. We compute the mean value and the standard deviation of the estimator of the quantity of interest employing standard (multifidelity) Monte Carlo estimators and our techniques, and then we plot the approximated Gaussian distributions of the estimators constructed from the approximated mean and variance. In order to take into account all the variance of the methods, we first get a pilot sample from the distributions of the input parameters, which we employ to get the best hyperparameters for the networks as outlined in \cref{rem:hyperparameters}, and to train them. Then, we compute the optimal allocation using the pilot sample and setting a computational budget of 300 high-fidelity simulations, and finally we discard the pilot sample and draw new samples from which we obtain the multifidelity estimation with a one-dimensional shared subspace. This procedure is repeated 100 times.

Let us now focus on the method in \cref{sec:MFMC_AS}. In \cref{fig:analytic_functions_AS} we consider four different cases, setting the size of the pilot sample equal to 100. In the first column we leverage the fact that we know the exact gradient of the functions \eqref{eq:exact_gradient_AS} and the exact normalizing flow from the input distributions \eqref{eq:exact_NF_AS}, and we plot the results in the ideal setting where we can employ them. In the second and third columns, we add one level of complexity at a time, by first computing the gradients through surrogate models based on fully connected neural networks, and then by training a spline-based normalizing flow. Finally, in the last column we consider the most general case, where we do not have any a priori knowledge of the gradients and the normalizing flow, which is actually employed in concrete applications. Both from the correlation values and the variances of the output distributions, we observe that in all four cases our approach outperforms standard (multifidelity) Monte Carlo techniques. In particular, we notice that the presence of the surrogate model seems not to affect the final results, and this is due to the fact that the functions representing the high-fidelity and low-fidelity models can be easily approximated by fully connected neural networks. We also note a slight increase in the variance when we replace the exact normalizing flow with its spline-based approximations, which, nevertheless, does not deteriorates the final output. Moreover, in \cref{fig:analytic_functions_AS_N} we compare the results varying the size of the pilot sample for the general case where we do not use the analytic gradients or the normalizing flow. We observe that the final variance of the estimator increases only when a significantly small pilot sample is drawn, meaning that in this case the estimator is not strongly sensitive to the number of data, and that it is not necessary to find the exact active subspace to achieve variance reduction, as long as the approximated important direction is not highly different from the real one.

We repeat similar experiments for the method in \cref{sec:MFMC_AE} with RealNVP as normalizing flow. In \cref{fig:analytic_functions_AE} we set the sample size equal to 100, and we consider four different cases. In the first two columns we fix the encoder to be the actual model as described in \cref{sec:particular_autoencoder}, while in the last two columns we study the general methodology where the autoencoder is trained using the available data. In both cases we use either the model itself or a surrogate model based on fully connected neural networks to compute the encoder or train the autoencoder, respectively. Similarly to the other approach, we observe that our technique is able to improve standard (multifidelity) Monte Carlo estimators, and that the presence of the surrogate model does not seem to affect the final results. Moreover, we do not notice a significant difference between the standard autoencoder approach and the one where the encoder is equal to the model. In \cref{fig:analytic_functions_AE_N} we also compare the results varying the size of the pilot sample for the last case where the autoencoder is trained using the surrogate model. We notice that the variance of the estimator is smaller when the number of data in the pilot sample is larger, i.e., if we have enough information to find the nonlinear subspace where the models vary the most. Finally, in \cref{fig:analytic_functions_corr} we show for a particular sample how the correlation increases from the original low-fidelity model to the new reduced low-fidelity models obtained applying our methodologies. It is interesting to notice that the autoencoder seems to introduce a nonnegligible bias, compared to the active subspace. Nevertheless, this is not important for the performance of the algorithm that only depends on the correlation between the models. We finally remark that, in order for the method with the autoencoder to be able to have a better performance with respect to the method with active subspaces, a larger number of data or a more complex problem is necessary, as we will see in the next examples.

\subsection{Reaction-diffusion equation}

\begin{figure}[t]
\centering
\begin{tabular}{ccccc}
\includegraphics{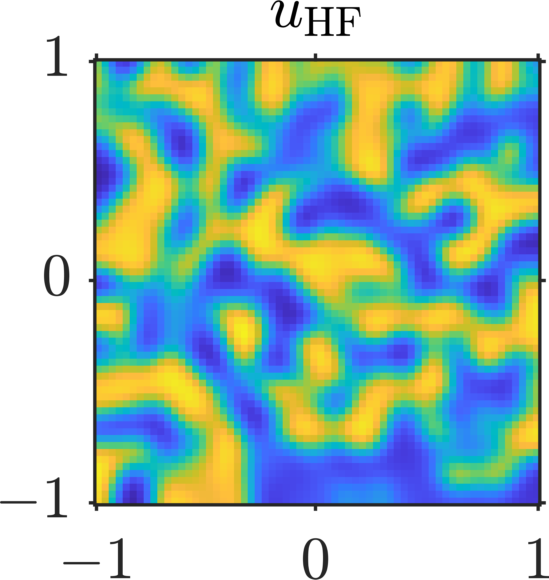} &
\includegraphics{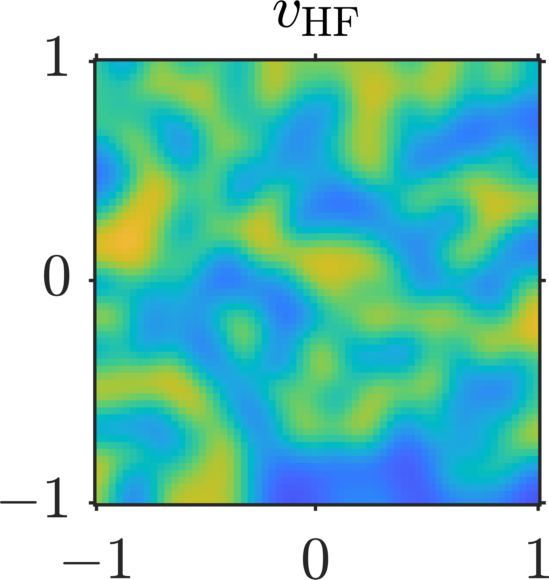} &
\includegraphics{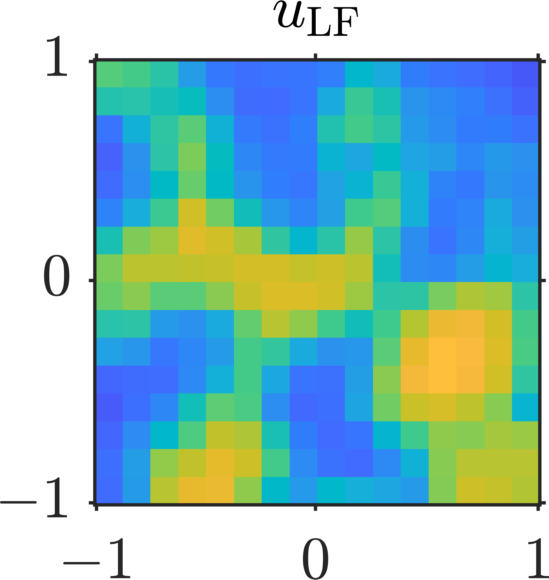} &
\includegraphics{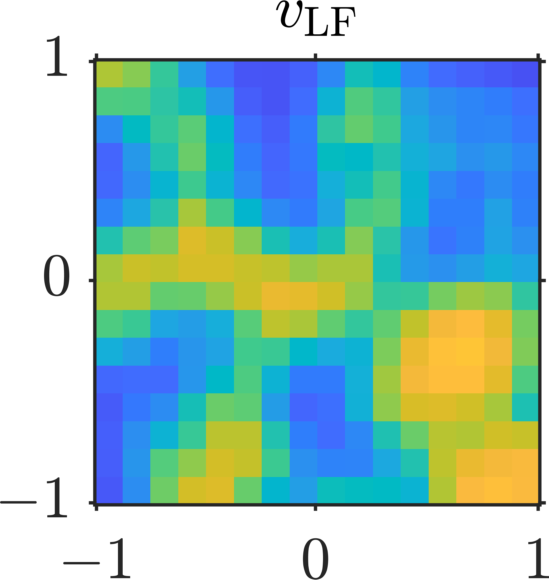} &
\includegraphics[scale=1.2]{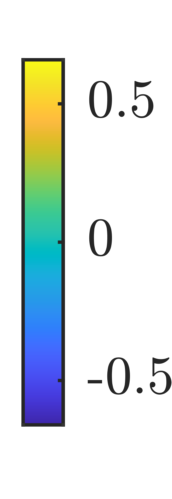}
\end{tabular}
\caption{High-fidelity and low-fidelity solutions of the reaction-diffusion equation at the final time $T=4$.}
\label{fig:PDE_solutions}
\end{figure}

\begin{figure}[t]
\centering
\begin{tabular}{cccc}
\includegraphics{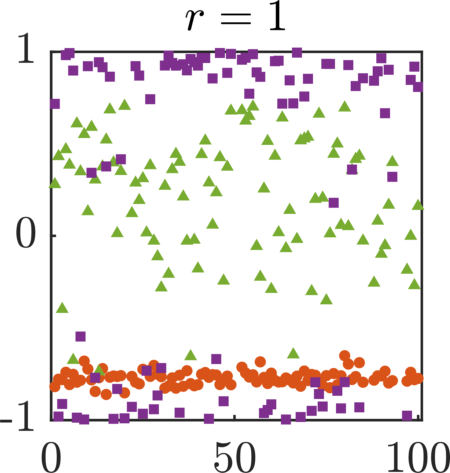} &
\includegraphics{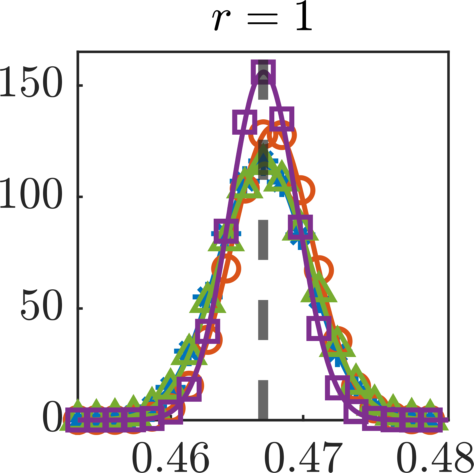} &
\includegraphics{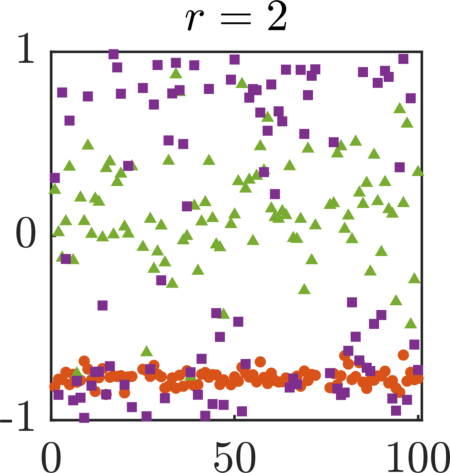} &
\includegraphics{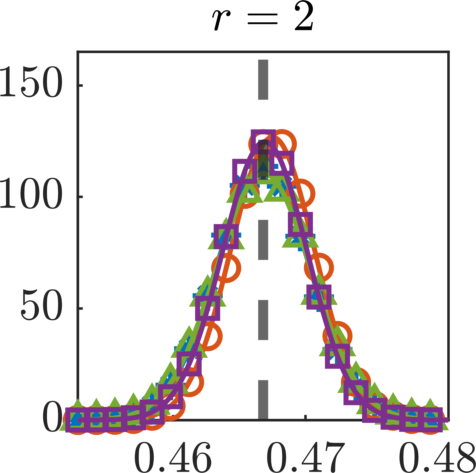}
\end{tabular}
\includegraphics[scale=0.2]{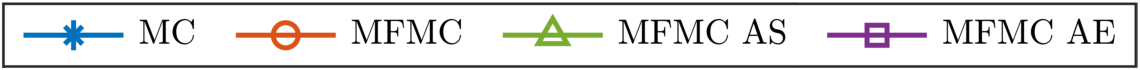}
\caption{Comparison between our methods (MFMC AS and MFMC AE) with standard (multifidelity) Monte Carlo (MC and MFMC) for the reaction-diffusion equation, varying the reduced dimension $r = 1,2$.}
\label{fig:PDE_MFMC}
\end{figure}

In this section we work with a more complex example taken from the PDEBench repository \cite{TPL23}, which has applications in real-world problems, i.e., biological pattern formation \cite{Tur52}. In particular, we consider the two-dimensional reaction-diffusion equation
\begin{equation} \label{eq:PDE_equations}
\begin{aligned}
\frac{\partial u(t,x,y)}{\partial t} &= D_u \frac{\partial^2 u(t,x,y)}{\partial x^2} + D_u \frac{\partial^2 u(t,x,y)}{\partial y^2} + R_u(u(t,x,y), v(t,x,y)), \\
\frac{\partial v(t,x,y)}{\partial t} &= D_v \frac{\partial^2 v(t,x,y)}{\partial x^2} + D_v \frac{\partial^2 v(t,x,y)}{\partial y^2} + R_v(u(t,x,y), v(t,x,y)),
\end{aligned}
\end{equation}
with no-flow Neumann boundary conditions
\begin{equation}
\begin{aligned}
D_u \frac{\partial u(t,x,y)}{\partial x} &= 0, \qquad D_u \frac{\partial u(t,x,y)}{\partial y} = 0, \qquad t \in (0,T], \quad x,y \in (-L,L), \\
D_v \frac{\partial v(t,x,y)}{\partial x} &= 0, \qquad D_v \frac{\partial v(t,x,y)}{\partial y} = 0, \qquad t \in (0,T], \quad x,y \in (-L,L),
\end{aligned}
\end{equation}
where $u,v \colon (0,T] \times (-L,L) \times (-L,L)$ are called the activator and the inhibitor, $T>0$ is the final time, $L>0$ is the size of the domain, $D_u,D_v > 0$ are the diffusion coefficients, and $R_u,R_v$ are the reaction functions
\begin{equation}
\begin{aligned}
R_u(u,v) &= u - u^3 - k - v, \\
R_v(u,v) &= u - v, 
\end{aligned}
\end{equation}
with $k>0$ and which are equivalent to the well-known Fitzhugh-Nagumo equations \cite{KlT84}. The initial condition is generated as standard normal random noise $u(0,x,y), v(0,x,y) \sim \mathcal N(0,1)$ for $x \in (-L,L)$ and $y \in (-L,L)$ only once and then fixed. We consider as our quantity of interest the sum of the averages of the absolute value of the activator and the inhibitor at the final time
\begin{equation}
Q = \frac1{4L^2} \int_{-L}^L \int_{-L}^L \abs{u(T,x,y)} \dd x \dd y + \frac1{4L^2} \int_{-L}^L \int_{-L}^L \abs{v(T,x,y)} \dd x \dd y,
\end{equation}
and we set the final time $T=4$ and the size of the domain $L=1$. The equations are solved employing the finite volume method for space discretization and the fourth order Runge--Kutta method for time integration. The high-fidelity model is obtained by setting 64 finite volume cells in each direction and running the simulation for 400 time steps, while for the low-fidelity model we use 16 finite volume cells in each direction and run for 100 time steps, which yields a cost ratio of $w = 0.1$. Moreover, in the low-fidelity model we replace the diffusion coefficients $D_u$ and $D_v$ by their average value, i.e., $\bar D = (D_u + D_v)/2$. The input parameters for the high-fidelity model are the diffusion coefficients $D_u, D_v$ and the reaction coefficient $k$, while for the low-fidelity model we have $\bar D$ and $k$, and therefore we are in the framework where the two models have a dissimilar parameterization. In \cref{fig:PDE_solutions} we plot the solutions $u$ and $v$ of equations \eqref{eq:PDE_equations} for both the high-fidelity and low-fidelity models, setting the parameters $D_u = 10^{-3}, D_v = 5 \cdot 10^{-3}, k = 10^{-3}$, and $\bar D = 3 \cdot 10^{-3}$. The input probability distribution for the forward uncertainty propagation study is such that $k$ is independent of $D_u$ and $D_v$, $k \sim \mathcal U([0.5 \cdot 10^{-3}, 1.5 \cdot 10^{-3}])$, and $D_u,D_v$ are uniformly distributed in the triangle with vertices $\{ (0.25 \cdot 10^{-3}, 4 \cdot 10^{-3}), (1.75 \cdot 10^{-3}, 5 \cdot 10^{-3}), (1 \cdot 10^{-3}, 6 \cdot 10^{-3}) \}$. We remark that this is an example of correlated random inputs, since $D_u$ and $D_v$ in the high-fidelity model are uniformly distributed on a triangular support, leading to correlated parameters. The distribution of $\bar D$ is then obtained accordingly. One of the challenges when working with dependent inputs relates to the difficulty of enforcing that the new low-fidelity samples generated by the pipeline are defined over the same support as the original low-fidelity inputs. To overcome this problem, we assume the low-fidelity model to be defined even outside the original support of input parameters, and we are currently investigating how to lift such assumption.

Similarly to the previous section, we compare our methods with standard (multifidelity) Monte Carlo techniques, by computing the mean value and the standard deviation of the estimators based on 100 samples, and plotting the corresponding Gaussian distribution. In the variance of the final distribution, we take into account all the possible sources of uncertainties, i.e., sampling from the input distribution and hyperparameter tuning. The optimal allocation is computed assuming a computational budget of 100 high-fidelity simulations.

The numerical results are displayed in \cref{fig:PDE_MFMC} for both one-dimensional and two-dimensional shared spaces. We observe that our method based on active subspaces (MFMC AS) is not able to reduce the variance of the estimator. This implies either that a linear transformation is not enough to capture the directions where the models vary the most, or that the surrogate models are not sufficiently accurate to provide a good approximation of the gradient of the models, and consequently of the active subspaces. Alternative techniques could be explored to compute gradients, such as ridge regression \cite{HoC18} or adaptive basis \cite{TiG14}. Nevertheless, we do not consider these approaches here since the approximation of the gradient is not the main focus of this work. On the other hand, our technique based on autoencoders, and therefore nonlinear transformations, outperforms all the other approaches and provides a significant variance reduction while increasing the correlation between the high-fidelity and low-fidelity models. Moreover, we notice that in this case a one-dimensional shared space yields better results than a two-dimensional subspace. Future work will focus on how to determine the optimal dimensionality of the reduced space. This problem is more challenging for the method based on autoencoders because we do not have an order of importance provided by the eigenvalues as in the active subspace technique.

\subsection{Cardiovascular simulation} \label{sec:cardio}

\begin{figure}[t]
\centering
\includegraphics[scale=0.1]{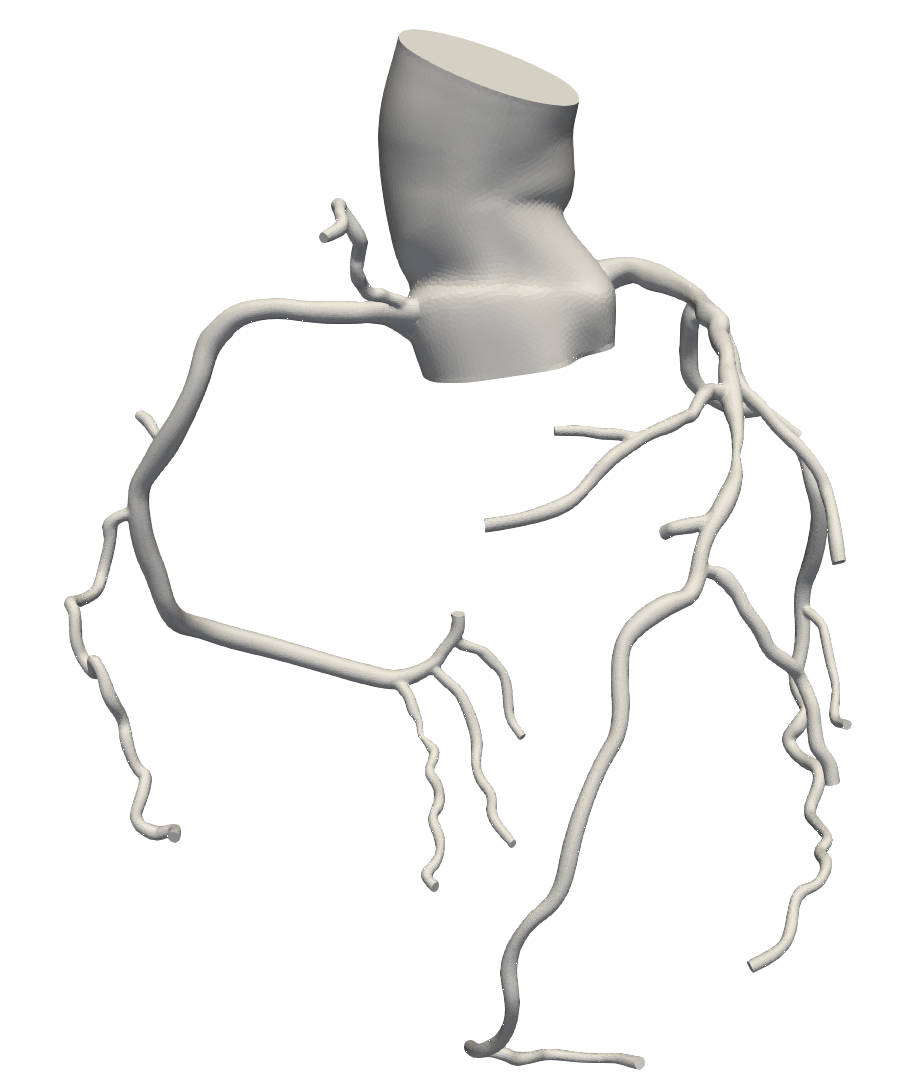}
\includegraphics{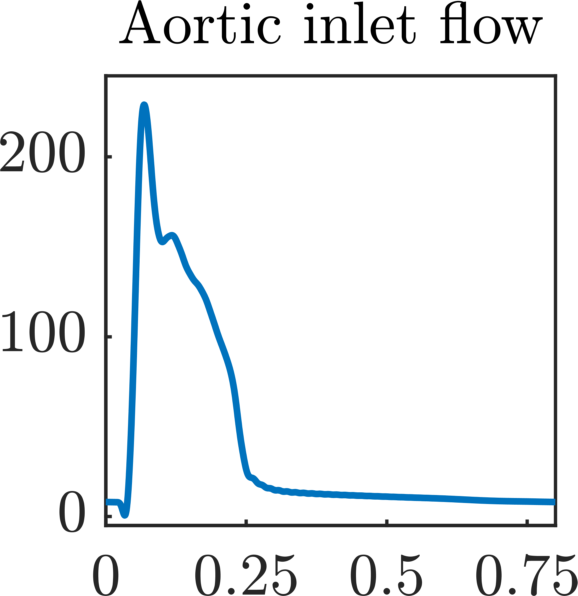}
\includegraphics{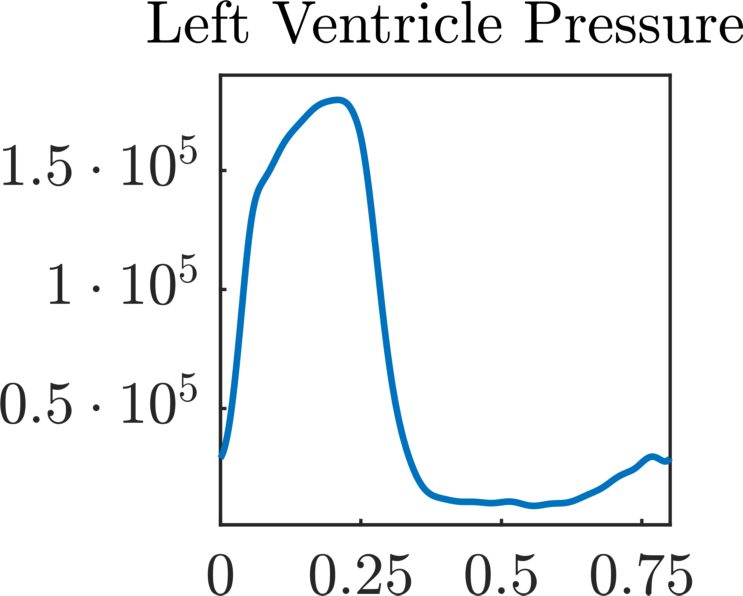}
\includegraphics{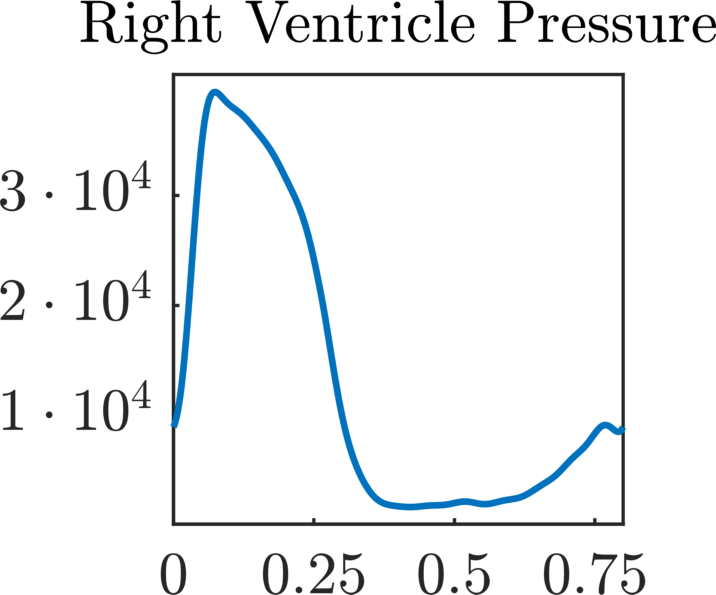}
\caption{Coronary model for the cardiovascular simulations, together with the prescribed profiles for aortic inlet flow, left ventricle pressure, and right ventricle pressure.}
\label{fig:cardiovascular_model}
\end{figure}

\begin{table}[t]
\centering
\begin{tabular}{cccc}
Parameter & Mean Value & Unit & Std Dev \\
\hline
Total resistance & $3193.5891$ & Barye/ml & $50 \%$ \\
Total aorta capacitance & $1.6877 \cdot 10^{-4}$ & ml/Barye & $30 \%$ \\
Total left coronary arteries capacitance & $7.2127 \cdot 10^{-6}$ & ml/Barye &  $30 \%$ \\
Total right coronary arteries capacitance & $7.2568 \cdot 10^{-6}$ & ml/Barye & $30 \%$ \\
Coronary Young's modulus & $11500000.0$ & Barye & $30 \%$ \\
Left coronary intramyocardial pressure & $1.405165$ & Barye & $20 \%$ \\
Right coronary intramyocardial pressure & $4.394061$ & Barye & $20 \%$
\end{tabular}
\caption{Input parameters and distribution of the cardiovascular simulations.}
\label{tab:cardiovascular_parameters}
\end{table}

\begin{figure}[t]
\centering
\includegraphics{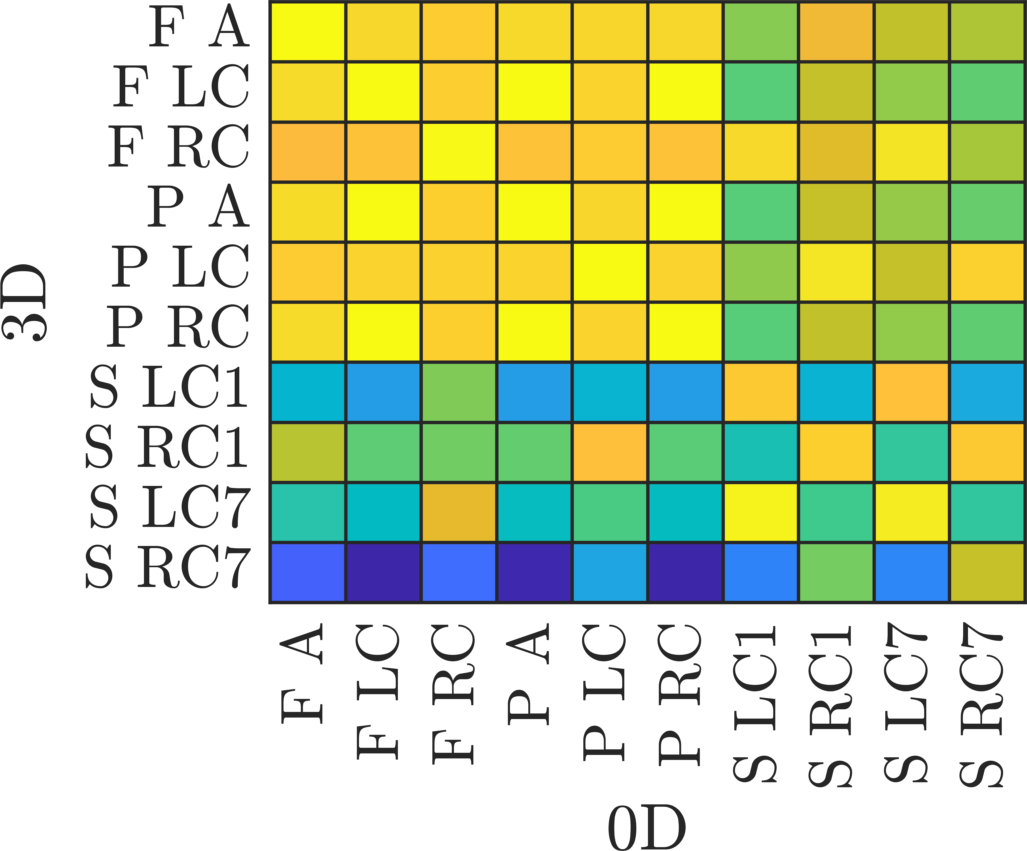} \hspace{0.25cm}
\includegraphics{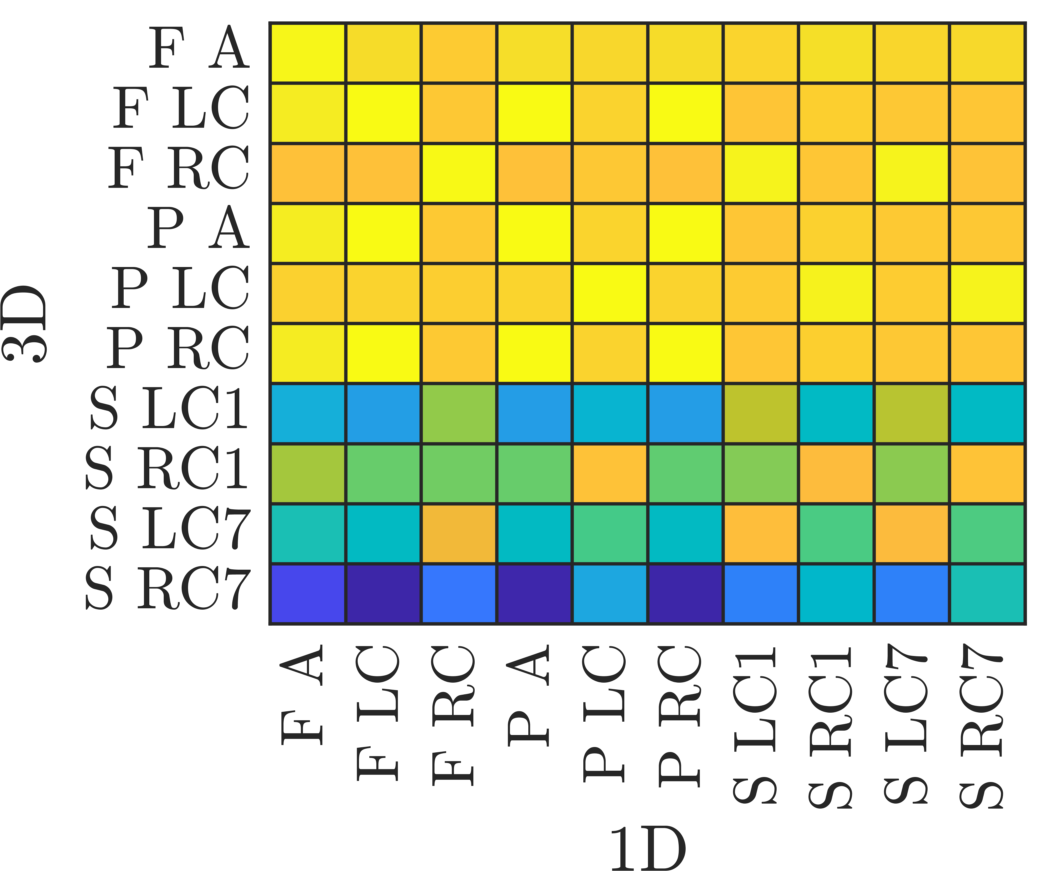} \hspace{0.5cm}
\includegraphics{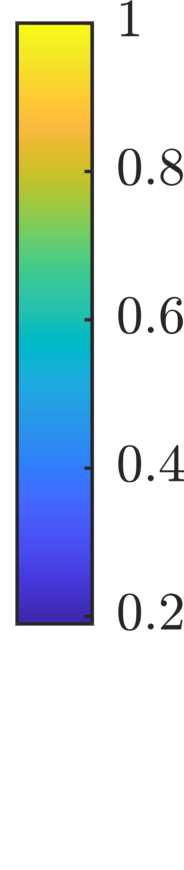} 
\caption{Matrix of the absolute value of the cross-correlation between 3D high-fidelity and 0D/1D low-fidelity cardiovascular simulations for the minimum value of different quantities of interest. The correlation between the same quantity of interest is represented on the diagonals. Notation: F flow, P pressure, S wall shear stress, A aorta, LC left coronary, RC right coronary.}
\label{fig:correlations}
\end{figure}

\begin{figure}[t]
\centering
\includegraphics{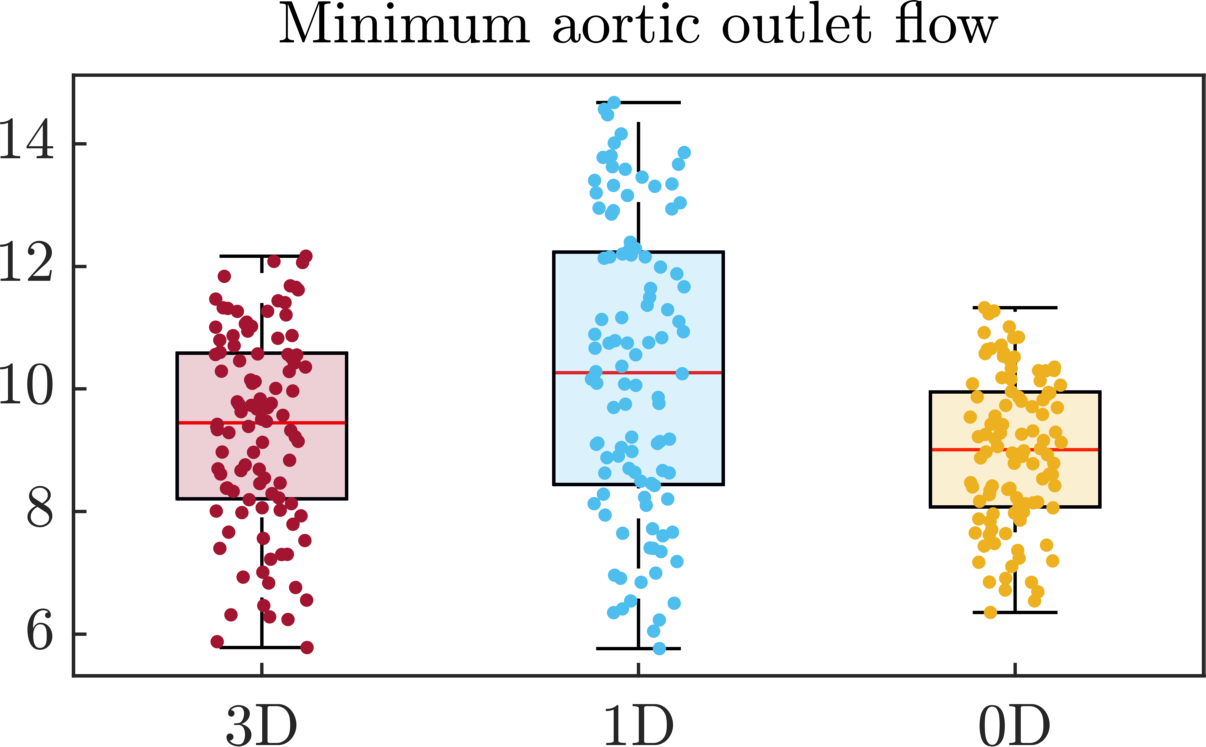} \hspace{1cm}
\includegraphics{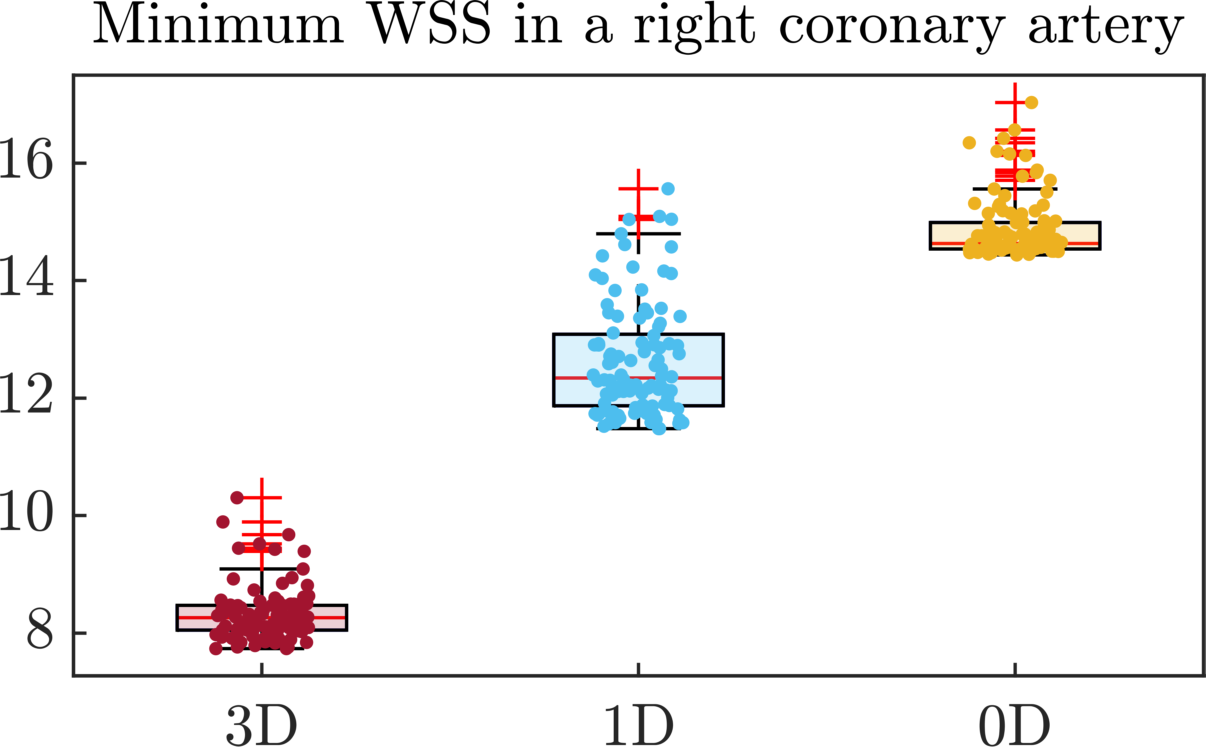}
\caption{Boxplots of the quantities of interest computed from the cardiovascular simulations, employing 3D, 1D, and 0D models. Left: Minimum aortic outlet flow. Right: minimum wall shear stress computed in the right coronary artery with a stenosis.}
\label{fig:cardiovascular_boxplot}
\end{figure}

\begin{table}[t]
\centering
\begin{tabular}{cccc}
&& 0D & 1D \\ \\
Same QOI &&
\begin{tabular}{ccc}
Method & $N^\HF$ & $N^\LF$ \\
\hline
MFMC & $103$ & $25721$ \\
MFMC AE & $103$ & $44371$
\end{tabular}
&
\begin{tabular}{ccc}
Method & $N^\HF$ & $N^\LF$ \\
\hline
MFMC & $101$ & $1870$ \\
MFMC AE & $99$ & $3246$
\end{tabular}
\\ \\
Different QOI &&
\begin{tabular}{ccc}
Method & $N^\HF$ & $N^\LF$ \\
\hline
MFMC & $104$ & $7133$ \\
MFMC AE & $103$ & $19297$
\end{tabular}
&
\begin{tabular}{ccc}
Method & $N^\HF$ & $N^\LF$ \\
\hline
MFMC & $103$ & $741$ \\
MFMC AE & $97$ & $4044$
\end{tabular}
\end{tabular}
\caption{Optimal allocation for the multifidelity estimators for the cardiovascular simulations. The quantity measured by the low-fidelity model, both 0D (left) and 1D (right), is either the same (top) or different (bottom) from the quantity of interest computed by the 3D high-fidelity model.}
\label{tab:cardiovascular_optimal_allocation}
\end{table}

\begin{table}[t]
\centering
\begin{tabular}{cccc}
&& 0D & 1D \\ \\
Same QOI &&
\begin{tabular}{ccc}
Method & Mean & Interval \\
\hline
MC & $8.332$ & $\pm 0.142$ \\
MFMC & $8.410$ & $\pm 0.079$ \\
MFMC AE & $8.381$ & $\pm 0.057$
\end{tabular}
&
\begin{tabular}{ccc}
Method & Mean & Interval \\
\hline
MC & $8.332$ & $\pm 0.142$ \\
MFMC & $8.332$ & $\pm 0.114$ \\
MFMC AE & $8.390$ & $\pm 0.089$
\end{tabular}
\\ \\
Different QOI &&
\begin{tabular}{ccc}
Method & Mean & Interval \\
\hline
MC & $8.332$ & $\pm 0.142$ \\
MFMC & $8.331$ & $\pm 0.136$ \\
MFMC AE & $8.435$ & $\pm 0.101$
\end{tabular}
&
\begin{tabular}{ccc}
Method & Mean & Interval \\
\hline
MC & $8.332$ & $\pm 0.142$ \\
MFMC & $8.339$ & $\pm 0.139$ \\
MFMC AE & $8.418$ & $\pm 0.079$
\end{tabular}
\end{tabular}
\caption{Values of the estimators (in Barye) together with their 90 \% confidence interval for the cardiovascular simulations. The quantity measured by the low-fidelity model, both 0D (left) and 1D (right), is either the same (top) or different (bottom) from the quantity of interest computed by the 3D high-fidelity model.}
\label{tab:cardiovascular_estimations}
\end{table}

\begin{figure}[t]
\centering
\begin{tabular}{cccc}
\includegraphics{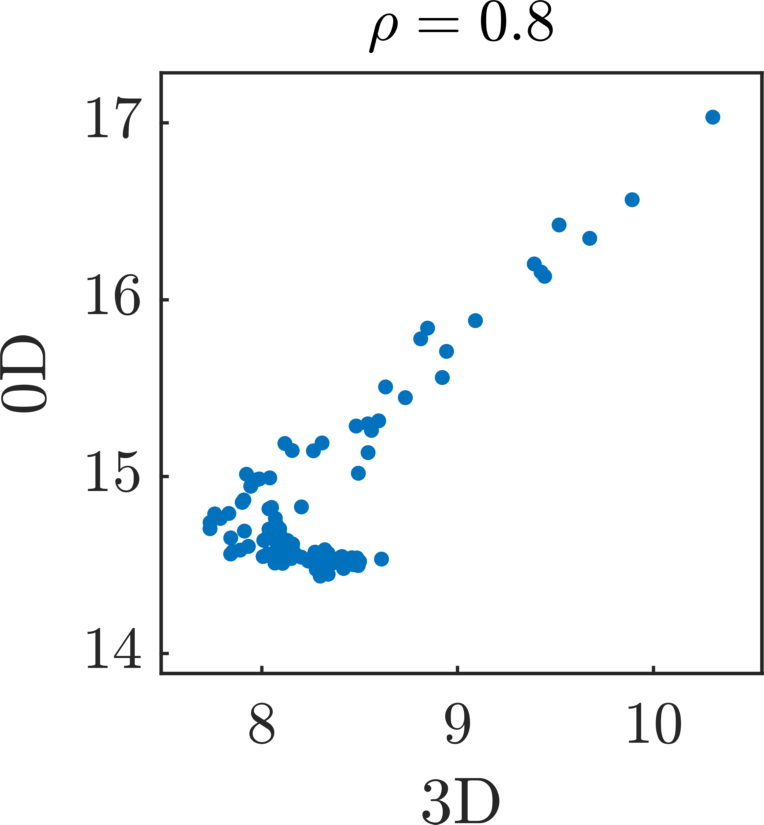} &
\includegraphics{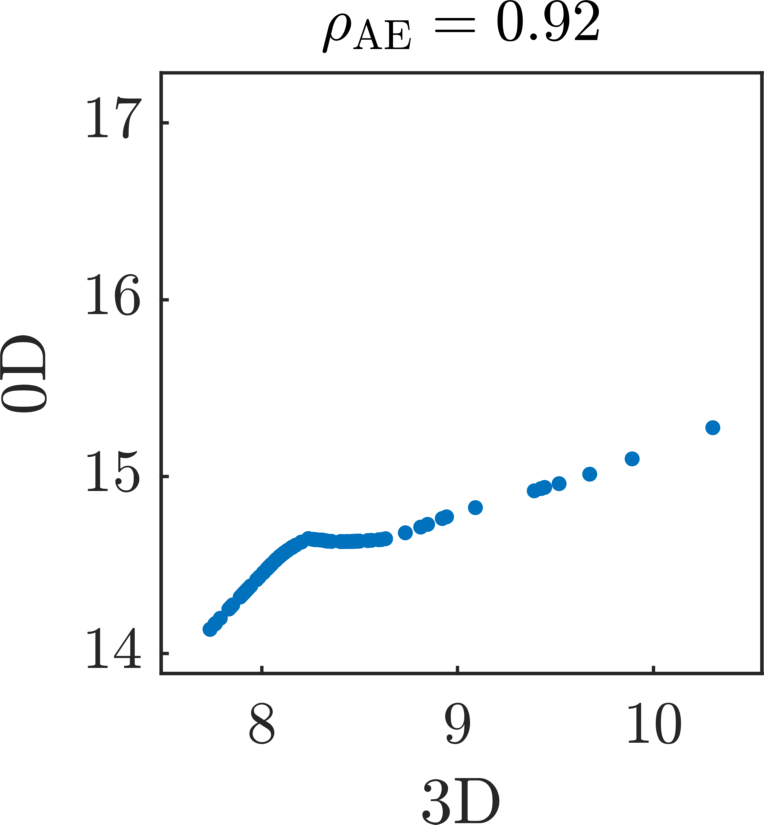} & $\quad$ &
\includegraphics{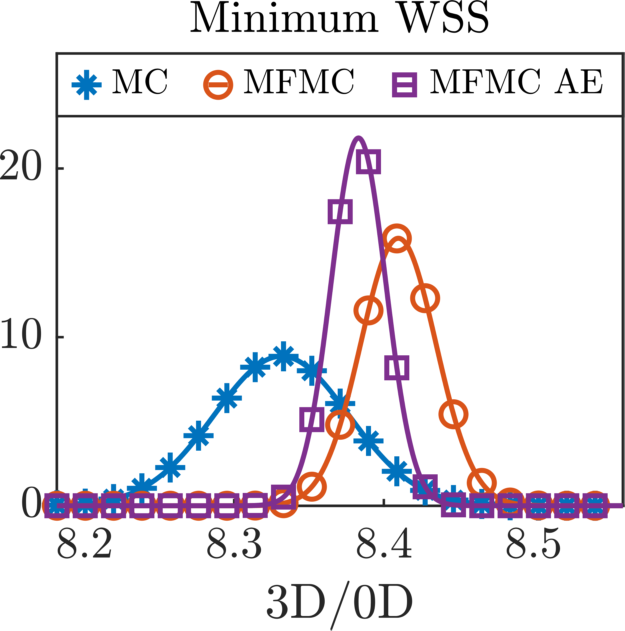} \\
\includegraphics{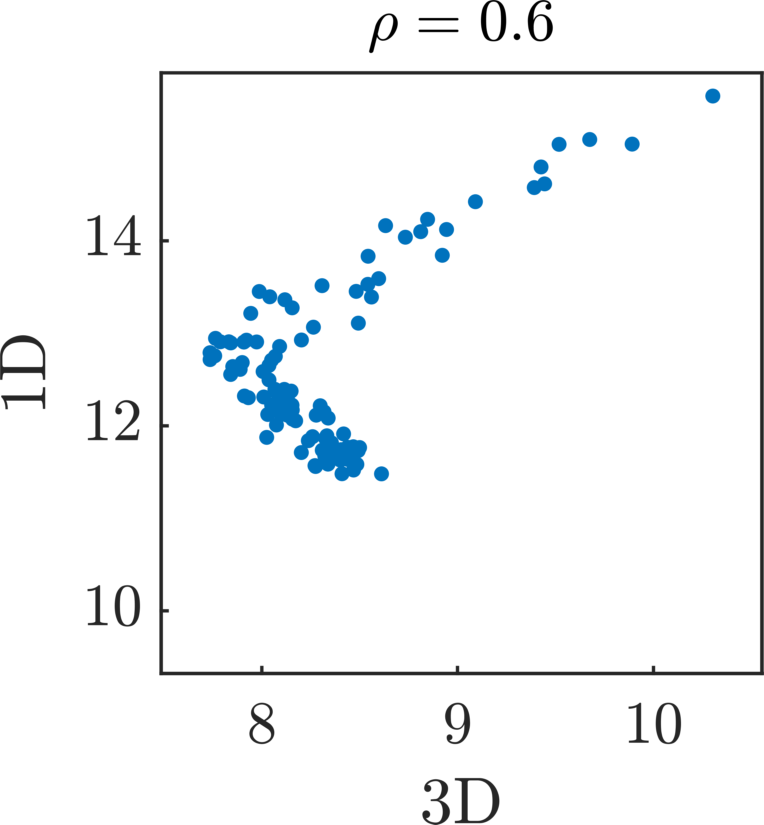} &
\includegraphics{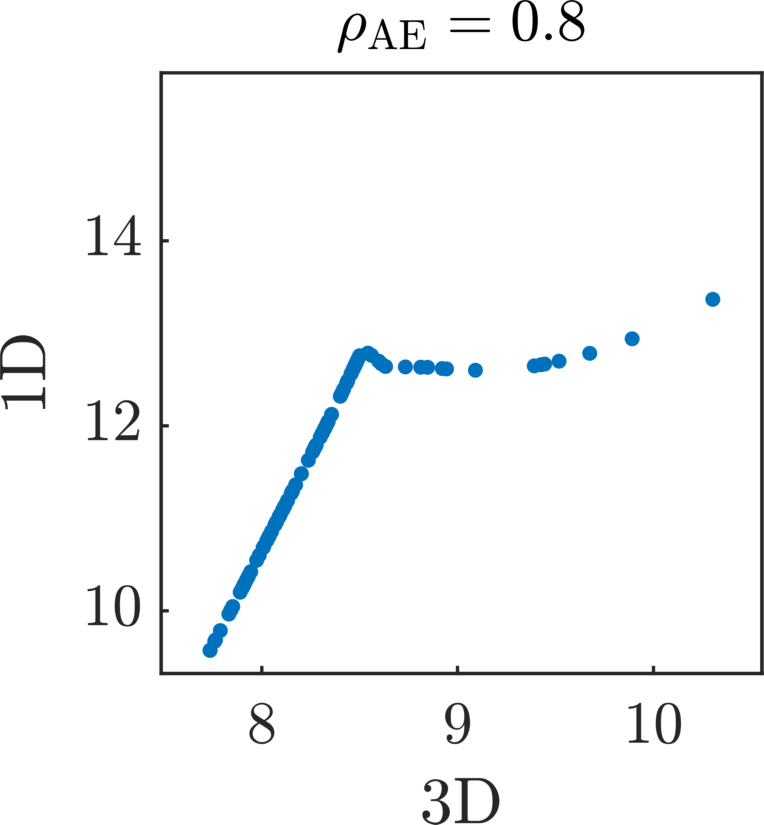} & $\quad$ &
\includegraphics{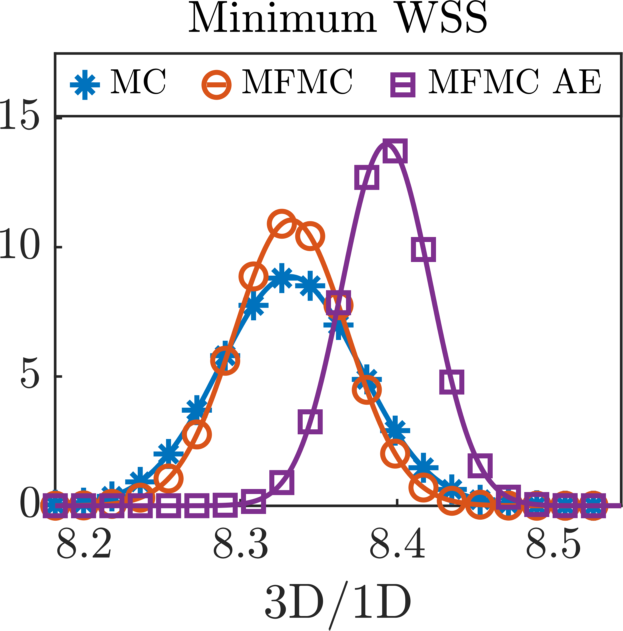}
\end{tabular}
\caption{Comparison between MFMC AE with standard (multifidelity) Monte Carlo (MC and MFMC) for the cardiovascular simulations, where the quantity estimated by the high-fidelity and low-fidelity model is the same (minimum wall shear stress in the right coronary artery with a stenosis). Top: 0D low-fidelity model. Bottom: 1D low-fidelity model. Left: correlation between high-fidelity (3D) and low-fidelity data employing MFMC and MFMC AE. Right: estimated density distribution for the quantity of interest.}
\label{fig:cardiovascular_same}
\end{figure}

\begin{figure}[t]
\centering
\begin{tabular}{cccc}
\includegraphics{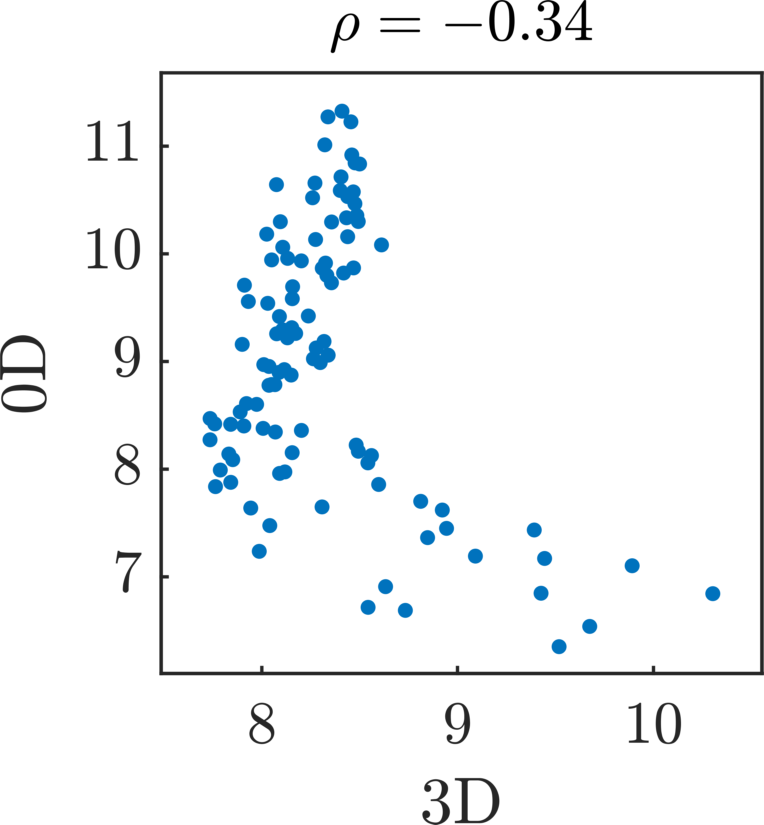} &
\includegraphics{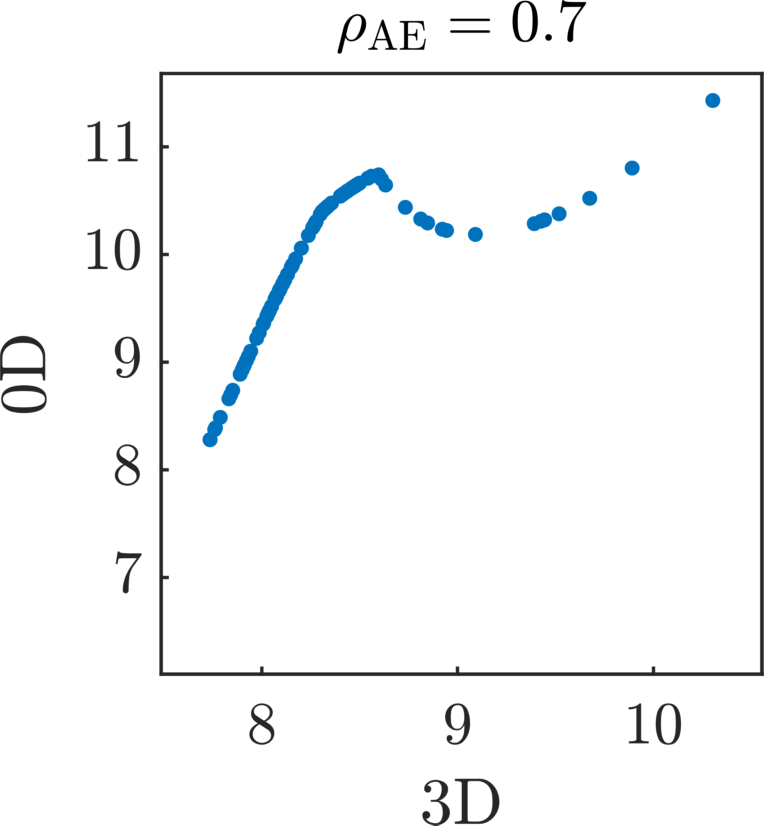} & $\quad$ &
\includegraphics{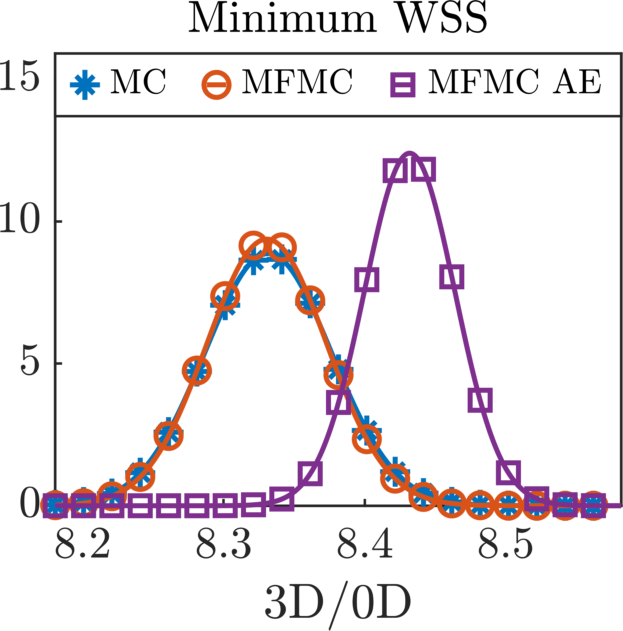} \\
\includegraphics{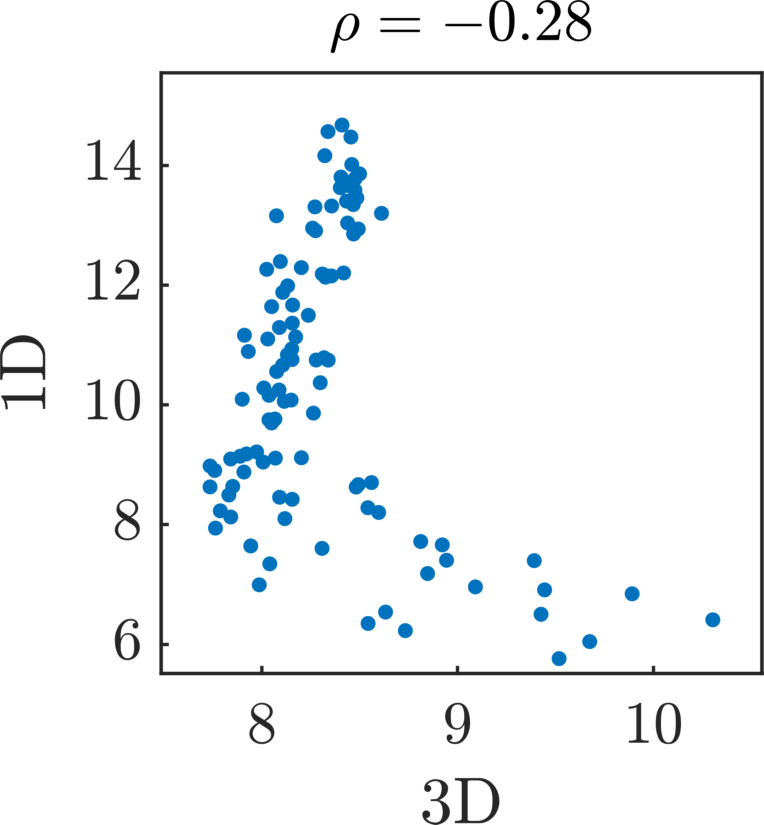} &
\includegraphics{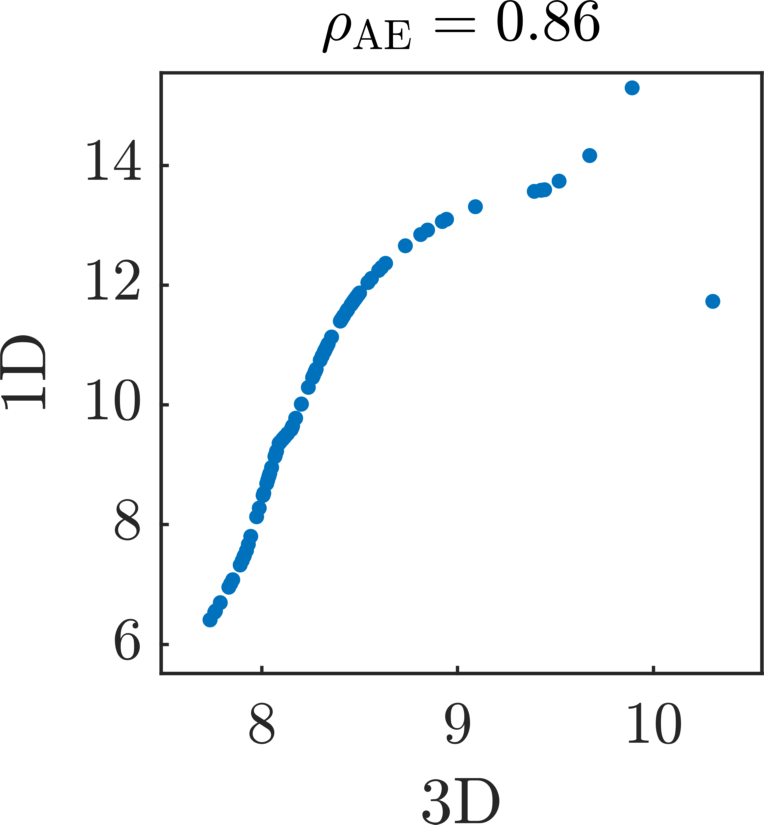} & $\quad$ &
\includegraphics{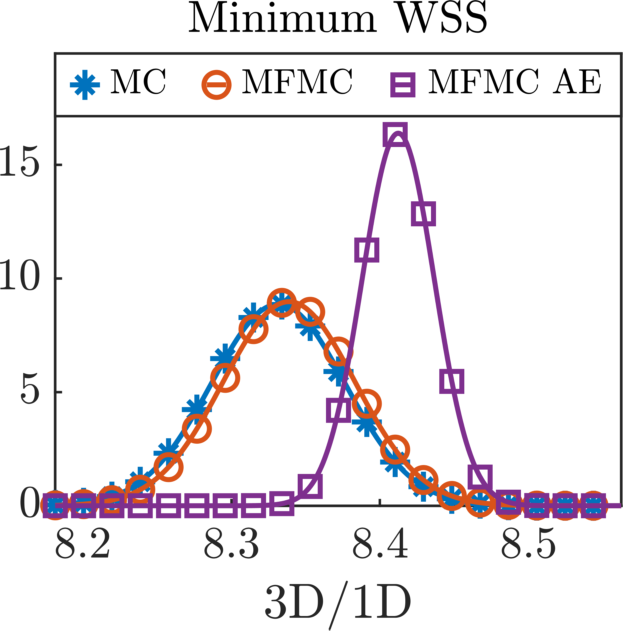}
\end{tabular}
\caption{Comparison between MFMC AE with standard (multifidelity) Monte Carlo (MC and MFMC) for the cardiovascular simulations, where the quantity estimated by the high-fidelity and low-fidelity model is different (minimum wall shear stress in the right coronary artery with a stenosis for the high-fidelity model and minimum aortic outlet flow for the low-fidelity model). Top: 0D low-fidelity model. Bottom: 1D low-fidelity model. Left: correlation between high-fidelity (3D) and low-fidelity data employing MFMC and MFMC AE. Right: estimated density distribution for the quantity of interest.}
\label{fig:cardiovascular_different}
\end{figure}

We now consider a real application for which we have a limited data availability. We focus on cardiovascular blood flow simulations of coronary artery disease. In \cref{fig:cardiovascular_model} we show the anatomic model together with the prescribed flow boundary condition at the aortic inlet and the intramyocardial pressures specified at the coronary outlets \cite{KVC10}. The high-fidelity model is represented by 3D simulations, for which we generate two different low-fidelity models: 1D and 0D simulations. One-dimensional hemodynamics models \cite{FNQ99} are formulated considering blood as a Newtonian fluid with velocity only along the axial direction of an ideal cylindrical branch, constant pressure over each vessel cross section, and a non-slip boundary condition at the vessel lumen. The governing equations are found by integrating the incompressible Navier-Stokes equations over the cross section of a deformable cylindrical domain, and the system of equations is completed with a constitutive model relating pressure to cross-sectional area deformations. On the other hand, zero-dimensional hemodynamics models are lumped parameter networks. They consist of an equivalent circuit model formulated by hydrodynamic analogy in terms of flow rate (electrical current) and pressure differences (voltage). Low fidelity models are created through an automated model generation pipeline recently implemented in SimVascular \cite{PPV22}. More details on the set up for the 3D simulations are provided in \cite{MSF23} and \cite{MKS23}.

The input parameters of the models are listed in \cref{tab:cardiovascular_parameters}, and each parameter is distributed as a uniform random variable in the interval \emph{Mean Value} $\cdot \; (1 \; \pm \; $\emph{Std Dev}$/100)$, independent of the others. These values have been tuned so that model outputs correspond to physiological patient data. We sample 104 different parameters from the input distribution, and perform the corresponding 3D, 1D, and 0D simulations. We remark that each 3D simulation with deformable walls took approximately 15 hours on the Sherlock HPC cluster at Stanford using 4 nodes with 24 cores on an AMD EPYC 7502 processor with 2.5 GHz base CPU clock time and 32 GB of memory. We assume the cost ratios between the models to be $w^{\mathrm{1D}} = 1.5 \cdot 10^{-3}$ and $w^{\mathrm{0D}} = 2.5 \cdot 10^{-5}$.

We initially analyze several quantities of interest: minimum/maximum/average values in time of the outlet flow and pressure profile in the aorta and both the left and right coronary arteries, and minimum/maximum/average in space of the time averaged wall shear stresses in each coronary artery. We observe that in most of the cases the correlation between the 3D high-fidelity model and the 1D/0D low-fidelity models is extremely high, in particular for flow and pressures values. This implies that standard multifidelity Monte Carlo is already able to provide a significant variance reduction with respect to the Monte Carlo estimator. On the other hand, for wall shear stresses the correlation is notably smaller and this justifies the application of our methodologies. Indeed, wall shear stresses, being quantities strictly related to the 3D domain, are challenging to approximate through reduced order models. Moreover, since our approaches, in contrast to standard multifidelity Monte Carlo, create a shared parameterization between the two fidelities, we also compute the cross-correlation between the minimum value of different quantities, meaning that the 1D/0D simulation can measure a different output than the 3D simulation. In \cref{fig:correlations} we plot the matrices of the absolute value of the cross-correlations for some quantities of interest, and observe that the correlation can be poor for some pairs. We remark that if in concrete applications there is available data for which the two fidelities are poorly correlated, then our approaches can improve the estimation.

We then select the minimum wall shear stress in the right coronary artery which has a stenosis (S RC7 in \cref{fig:correlations}) as the quantity of interest to approximate for all the 3D, 1D, and 0D results. Moreover, to showcase the versatility of our methodologies, we also consider a different quantity of interest for the low-fidelity model with respect to the high-fidelity. This opens the possibility to employ multifidelity estimation for any available data independently of the fact that they are related to the quantity under investigation. In particular, we then extract the minimum aortic outlet flow (F A in \cref{fig:correlations}) as quantity of interest for the 1D and 0D simulations, which is readily available. In \cref{fig:cardiovascular_boxplot} we show the boxplots of the two quantities under consideration in this study. We observe that the aortic outlet flow can be better approximated by low-fidelity models compared to the wall shear stress.

We remark that, given the small number of simulations for 7-dimensional data, the standard active subspace approach introduced in this work does not provide better results than standard (multifidelity) Monte Carlo estimators. We therefore focus here on the particular choice of autoencoder described in \cref{sec:particular_autoencoder}, which only requires  training a fully connected neural network surrogate for the 3D simulations, and two one-dimensional normalizing flows. The optimal allocation problems are solved assuming a computational budget of 104 simulations, in order to be able to compare the results with the standard Monte Carlo approach, and the resulting number of high-fidelity and low-fidelity simulations are rounded to the closest integer and given in \cref{tab:cardiovascular_optimal_allocation}. In \cref{fig:cardiovascular_same} we show the results for the case when the quantity of interest computed by the low-fidelity model is the same quantity computed by the high-fidelity model, and in \cref{fig:cardiovascular_different} when it is different. Moreover, in \cref{tab:cardiovascular_estimations} we give the values with a 90 \% confidence interval of the realizations of the estimators. The intervals are obtained by multiplying the standard deviation given by equation \eqref{eq:variance_MFMC} by $\sqrt{10}$ since due to the Chebyshev's inequality we have
\begin{equation}
\Pr( |\widehat Q - \E[Q(\bm \xi)] \le \sqrt{10} \sigma) \ge 0.9,
\end{equation}
where $\widehat Q$ stands for any estimator and $\sigma$ for its standard deviation. In both cases we first notice that the correlations of the new reduced low-fidelity models increase, and this implies a reduction in the variance of the estimators. The probability distributions in the last columns are indeed Gaussian distributions with mean given by the value of the estimator and variance computed from equation \eqref{eq:variance_MFMC}. We remark that the difference in estimation between multifidelity estimators with respect to standard Monte Carlo observed in the last column does not correspond to any bias. In fact, we just compute a single realization of the estimators and give a visual representation of these values, which are also reported in \cref{tab:cardiovascular_estimations}. In particular, these plots have to be interpreted differently from the similar plots in the previous sections. All the estimators are indeed asymptotically unbiased since we did not modify the high-fidelity model in their definition, and, if we could repeat the experiments multiple times, the average of the estimated values would give the exact mean of the quantity of interest, i.e., the minimum wall shear stress in the coronary artery with the stenosis. We finally notice that in \cref{fig:cardiovascular_different} the fact that the quantities of interest of the low-fidelity and high-fidelity models are different does not affect the final results.

\section{Conclusions} \label{sec:conclusion}

In this work we proposed two different methodologies to improve multifidelity estimators for uncertainty propagation. In particular, we achieved variance reduction of the standard multifidelity Monte Carlo estimator by modifying the low-fidelity model in order to increase the correlation with the high-fidelity model. Our approaches rely on a shared space where the models vary the most and the parameters are distributed according to a standard Gaussian. We constructed the shared space through either linear or nonlinear dimensionality reduction techniques, namely active subspaces and autoencoders. We demonstrated by means of numerical experiments that, given sufficient data, the latter are able to find nonlinear transformations which allow us to further decrease the variance of the estimator with respect to linear transformations. Moreover, we employed normalizing flows to map different probability distributions into the same distribution, i.e., a standard Gaussian, and therefore generate a shared space. Our techniques not only permit getting an estimator with reduced variance, but also increase the range of applicability of multifidelity estimators. In particular, we allow for models with dissimilar parameterization, meaning that the number and type of input parameters between the high-fidelity and low-fidelity models and their distributions can be different. This implies that these approaches can also be applied to models which measure different quantities of interest which are not directly related, as long as the high-fidelity and the modified low-fidelity models are well correlated, as we showed in the challenging numerical examples involving cardiovascular simulations. A limitation of our approach is that a large amount of data might be necessary to train the surrogate models, find the lower-dimensional subspaces, and build the normalizing flows. However, as we showed in \cref{fig:analytic_functions_AS_N,fig:analytic_functions_AE_N} where we varied the number of data points, even if the best lower-dimensional manifolds are not correctly identified, we still have an improvement in terms of variance with respect to standard multifidelity Monte Carlo estimators. In other words, through the examples in the paper, we numerically show that the number of samples needed to build an accurate surrogate is typically much larger than those needed to build a surrogate for the sole purpose of identifying a shared space leading to a smaller variance than multifidelity Monte Carlo. The sensitivity analysis with respect to the number of samples also shows that, as expected, the variance of the resulting estimators decreases with a larger amount of data. Moreover, the construction of an accurate surrogate model could be lifted in the linear dimension reduction case by adopting strategies like adaptive basis (see, e.g.,~\cite{ZGE23}) or it could be directly learned on the latent space together with the autoencoder. We leave this latter approach, which we expect to require a significantly smaller amount of data due to the reduced dimension, for future work. Another interesting extension of the current method is to train the autoencoders of both the high-fidelity and low-fidelity models together, and include a term in the loss function that maximizes the resulting correlation and therefore improves the variance of the estimators. Finally, we are also interested in studying a method to automatically find the optimal reduced dimension of the shared space, which is fundamental for applications with high-dimensional input parameters.

\subsection*{Acknowledgements} 

This work is supported by NSF CAREER award \#1942662 (DES), NSF CDS\&E award \#2104831 (DES), NSF award \#2105345 (ALM), and NIH grant \#5R01HL141712 (ALM, KM). This work used computational resources from the Stanford Research Computing Center (SRCC). Sandia National Laboratories is a multi-mission laboratory managed and operated by National Technology \& Engineering Solutions of Sandia, LLC (NTESS), a wholly owned subsidiary of Honeywell International Inc., for the U.S. Department of Energy’s National Nuclear Security Administration (DOE/NNSA) under contract DE-NA0003525. This written work is authored by an employee of NTESS. The employee, not NTESS, owns the right, title and interest in and to the written work and is responsible for its contents. Any subjective views or opinions that might be expressed in the written work do not necessarily represent the views of the U.S. Government. The publisher acknowledges that the U.S. Government retains a non-exclusive, paid-up, irrevocable, world-wide license to publish or reproduce the published form of this written work or allow others to do so, for U.S. Government purposes. The DOE will provide public access to results of federally sponsored research in accordance with the DOE Public Access Plan. The authors thank Boris Kramer for insightful suggestions about Section 3.3.

\bibliographystyle{siamnodash}
\bibliography{biblio}

\end{document}